\documentstyle[preprint,aps,eqsecnum]{revtex}
\begin{document}
\draft
\tightenlines

\title{Effective ATI channels in high harmonic generation}

\author{M.~Yu.~Kuchiev and V.~N.~Ostrovsky~\cite{SP}}

\address{School of Physics, University of New South Wales,
Sydney 2052, Australia}

\maketitle

\begin{abstract}
Harmonic generation by an atom in a laser field is described
by the three-step mechanism as proceeding via above-threshold 
ionization (ATI) followed by the electron propagation in the 
laser-dressed continuum and the subsequent laser assisted 
recombination (LAR). An amplitude of harmonic production is 
given by the coherent sum of contributions from different 
intermediate ATI channels labeled by the number $m$ of 
absorbed laser photons. The range of $m$-values that gives 
substantial contribution is explored and found to be rather broad
for high harmonic generation. The coherence effects are
of crucial importance being responsible for the characteristic
pattern of harmonic intensities with a {\it plateau}\/ domain
followed by a {\it cutoff}\/ region. Due to multiphoton nature
of the process, an efficient summation of $m$-contributions 
can be carried out in the framework of  
the saddle point method. The saddle points correspond to some 
complex-valued labels $m=m_c$ associated with the intermediate 
{\it effective ATI channels}\/ in the three-step harmonic 
generation process. The advantage of this approach stems
from the fact that summation over large number of conventional 
ATI $m$-channels is replaced by summation over small number 
of effective $m_c$-channels. The equation governing $m_c$ has a 
transparent physical meaning: the electron ejected from 
the atom on the first (ATI) stage should return to the core 
to make LAR possible. The effective channel labels
$m$ move along characteristic trajectories  
in the complex plane as the system parameters vary.
In the cutoff region of the harmonic spectrum a single effective channel
contributes. For lower harmonics, in the plateau domain, 
two effective ATI channels become essential.
The interference of their contributions leads to oscillatory
pattern in the harmonic generation rates.
The calculated rates are in good agreement with
the results obtained by other approaches.

\end{abstract}


%\twocolumn
%\narrowtext

\section{INTRODUCTION}\label{Int}

\subsection{Three-step mechanism of HHG} \label{int1}

High efficiency of various processes in strong laser field could be understood basing on the observation that
the field-induced {\it quiver motion}\/ supplies an electron 
with high instantaneous energy.
Rescattering of the energetic electron on atomic core
generally is accompanied by the energy exchange between
electron, core and electromagnetic field.
In particular, the core could be excited or ionized
[double ionization (DI) of an atom]
and the high-energy photons could be emitted.
The latter process is known as the harmonic generation (HG). 
If one considers the active electron initially bound to the core,
then the electron at first should be (virtually) released
to the quasi-free state, as a precondition that the subsequent
events listed above become possible. As a whole, this
constitutes {\it three-step mechanism}\/
comprising above-threshold ionization (ATI),
propagation in laser-dressed continuum and the final
step which is electron-atom impact in case of rescattering,
core excitation or ionization,
or laser assisted recombination (LAR) in case of HG process.
Strong interaction between the receding
electron and the core is omitted in the standard
Keldysh \cite{Keldysh} model of multiphoton ionization.  
Its importance was first pointed out by Kuchiev \cite{Ku87}, 
who had predicted several phenomena where the electron-core 
interaction plays a crucial role. The related 
mechanism was named {\it ``atomic antenna''} to stress
the role of active electron in gaining energy from laser field.

Specifically for HG the three-step model was promoted in
the hybrid classical-quantum framework by Corkum \cite{C},
see also the papers by Kulander {\it et al}\/ 
\cite{KulanderA,KulanderB}. The subsequent theoretical 
developments were based on more sophisticated approaches 
and led to important advancements 
\cite{Hu,Lew,Lewphase,Bothers,Brapid,B},
albeit the three-step nature of the HG process was somewhat
veiled in these formulations.
Most clearly the three-step mechanism is exposed
in the framework of {\it factorization}\/ method
developed in Ref.~\cite{K95,K96} as an implementation of 
idea of atomic antenna of Ref.~\cite{Ku87} and applied to DI
process. This technique allows one to present quantum amplitude  
as a direct sum over contributions of intermediate ATI channels.
The application to HG process worked out by Kuchiev and Ostrovsky
\cite{KOlet,KOpap} quantitatively demonstrated validity and 
power of this approach. Since this development is the starting 
point of the present study, we briefly summarize it in 
Sec.~\ref{threestep} below.

The intermediate ATI channels form a discrete set being labeled 
by a number $m$ of absorbed photons. Generally speaking,
entire manifold of $m$-channels contributes to the rate of 
three-step process under consideration. One can anticipate that 
actually only some effective $m$-range should be essential, 
but this issue has not been investigated previously.
Another important question arises from the fact that
contributions to the rate of three-step process
from different ATI channels are to be summed coherently.
How essential are {\it coherence effects}\/ in reality remained
unclear. The acuteness of this problem is enhanced
in view of recent publications by Becker and Faisal 
\cite{Faisal1,Faisal2,BF} where semiempirical formula with 
{\it incoherent}\/ summation is suggested
in application to the DI process.

In the present paper we address these problems considering HG 
as the most simple three-step event. Being sharply important 
in itself, it hopefully provides a useful testground for other, 
more complicated processes. Our study shows that the range of 
efficiently contributing $m$-channels is rather broad 
(Sec.~\ref{phen}). It is defined by interplay of two factors
characterizing each $m$-channel contribution: the well known
ATI amplitude (Appendix \ref{A1}) and much less investigated 
LAR amplitude (Appendix \ref{LAR}). Importantly, the phases 
of these amplitudes play a major role, showing that 
the coherence effects are crucial.
In particular, solely these effects are responsible for the
well-known rapid fall-off of HG rates beyond the so called
{\it plateau}\/ domain. At once this finding provokes a new question:
how one can simplify summation over broad range of essential
intermediate $m$-channels. We achieve this objective (Sec.~\ref{eff})
by introducing a concept of {\it effective ATI channels},
or more briefly, effective channels (EC). Each EC is characterized 
by a complex-valued channel label $m_c$, i.e. a complex-valued
number of absorbed photons. The latter is defined from the
equation with simple and appealing physical meaning. 
The summation over a large number of intermediate $m$-channels is 
replaced by summation over very small number of EC (usually one or two). 
In the higher part of harmonic spectrum, beyond cutoff, it is 
sufficient to take into account a single EC. Two EC provide 
description of the plateau domain, including intricate oscillatory 
interference pattern of HG rates. 
As shown in Sec.~\ref{app}, EC approximation ensures good 
quantitative agreement with the harmonic intensities obtained 
within other approaches \cite{B,KOlet,KOpap}.

As the system parameters vary, the EC labels $m_c$ move along
characteristic trajectories in the complex-$m$ plane. 
This pattern condensedly expresses theoretical background 
of HG process in various regimes. For instance, transition 
between the regimes of one and two EC, i.e., between the 
cutoff and plateau domains, corresponds to ``collision'' 
of two trajectories in the complex plane. The oscillatory
behaviour of HG rate in the plateau domain is explained
by interference of two ECs. Thus ECs provide a unified 
framework for qualitative assessment of three-step
processes as well as for their quantitative description.
Probably the nearest analogy for this theoretical tool
comprise the well-known Regge poles \cite{Regge}, i.e., 
the states with complex-valued angular momentum, that give 
an effective description in the quantum scattering theory.

\subsection{Direct implementation of three-step mechanism in 
theory}
\label{threestep}

Now it is worthwhile to look in more detail how the three-step 
mechanism of HG is implemented in the theory of 
Refs.~\cite{KOlet,KOpap}. The key formula for the amplitude
of $N$-th harmonic generation,
\begin{mathletters} \label{dfin}
\begin{eqnarray} \label{sumc} 
d^+_N & = & 2 \, \sum_m \, d^+_{Nm} ~,
\\ \label{dc}
d^+_{Nm} & = & A_{m \, \mu_0}({\bf K}_m) 
\, B_{N \, m \mu_0}({\bf K}_m) ~,
\end{eqnarray}
\end{mathletters}
presents it as a sum where each term is a product of two
amplitudes of {\it physical, fully accomplished and observable}\/
processes; no ``off-shell'' entities appear.
The {\it first step}\/ is described by the first factor 
$ A_{m \, \mu_0}({\bf K}_m)$ which is an amplitude of physical 
ATI process when after absorption of $m$ laser photons
the active electron acquires a translational momentum 
${\bf p} = {\bf K}_m$; see detailed description of this amplitude
within the Keldysh-type approach in Appendix \ref{A1}. 
The other factor, $B_{N \, m \mu_0}({\bf K}_m)$, is a combined
amplitude of the {\it second and third steps}, i.e.,
propagation and laser assisted recombination (PLAR) 
amplitude. Under an additional approximation it can be 
factorized into propagation (expansion) factor 
$1/R_{m \mu_0}$ 
describing the {\it second step}\/ and the amplitude of the 
{\it third step}, LAR, $C_{N m}({\bf K}_m)$:
\begin{eqnarray} \label{prfact}
B_{N \, m \mu_0}({\bf K}_m)  = \frac{1}{R_{m \mu_0}} \, 
C_{N m}({\bf K}_m) ~.
\end{eqnarray}
$R_{m \mu_0}$ is merely an approximate expression for the distance
passed by the active electron in course of its laser-induced
wiggling motion, see Refs.~\cite{K95,KOlet,KOpap} and formula 
(\ref{propap}) below. The amplitude $C_{N m}({\bf K}_m)$ 
of the physical LAR process describes recombination, i.e.,
transition of electron with momentum ${\bf K}_m$ from 
the continuum to bound state. Since the continuum state is 
laser-dressed, the recombining electron can emit the $N$-th 
harmonic photon, gaining necessary extra energy from 
the laser field. The formulas for LAR and PLAR
amplitudes can be found in the Appendix \ref{A2}.

The summation in formula (\ref{dfin}) runs over a number of photons 
$m$ absorbed on the first step. Thus in the laser-dressed 
continuum the energy conservation constraint selects
the discrete set of ATI channels, where the electron has 
a translational momentum ${\bf K}_m$. These channels serve 
as intermediate states for the three-step HG process. 
To specify exactly, the {\it absolute value}\/ of the electron 
momentum in $m$-th ATI channel is defined by the energy 
conservation constraint in ATI as 
%\begin{eqnarray} \label{ec}
%\frac{1}{2} p^2 = m \omega - U_p + E_a ~,
%\end{eqnarray} 
\begin{equation}\label{Kmnm}
K_{m} =\sqrt{ 2 \left( m \omega - U_p
%\frac{e^2 F^2 }{4 \omega^2} 
+ E_a \right) } ~.
\end{equation}
where $E_a = - \frac{1}{2} \kappa^2$ is electron energy in
the initial bound state, $U_p \equiv F^2/(4 \omega^2)$ is 
the well-known ponderomotive potential, $\omega = 2 \pi / T$ 
is the laser frequency, $T$ is the period,
${\bf F}$ is the electric vector in the linear-polarized laser wave.
ATI can play role of the first stage of HG process only if
the electron momentum has specific {\it direction},
namely ${\bf K}_m$ is directed along ${\bf F}$.
This ensures eventual electron return to the core that makes
the final step, LAR, possible as discussed in detail in
Refs.~\cite{K95,KOlet,KOpap}.
It is worthwhile to indicate once again that our approach presumes 
the single active electron approximation; the atomic units are 
used throughout the paper unless indicated otherwise. 
The observable HG rates ${\cal R}_N$ are expressed via the amplitudes 
as 
\begin{eqnarray} \label{RN}
{\cal R}_N \equiv \frac{\omega^3 N^3}{2 \pi c^3} \, 
\left| d_N^+ \right|^2 ~,
\end{eqnarray}
$\Omega=N\omega$ is the frequency of emitted harmonic, 
$c$ is the velocity of light.

As already mentioned, the general framework for three-step 
decomposition of complicated laser-induced processes is provided
by factorization technique of Ref.~\cite{K95}. The accuracy of
this theoretical device is governed by multiphoton nature of
the processes: the larger is number of laser quanta involved,
the more accurate results it provides. To avoid confusion, it
should be emphasized that the factorization {\it is not}\/
related to the conventional perturbation theory, or to its
simplified version known as the pole approximation.

Here we only briefly outline some features of the derivation 
\cite{K95,KOlet,KOpap}
necessary for understanding of the present development.
The basic expression for the HG amplitudes contains
integration over two time variables, $t$ and $t^\prime$.
The factorization technique allowed us to disentangle these
integrations basing on the adiabatic approach.
Mathematically the latter implies the saddle point 
approximation for calculating the integrals which emerge
when a particular representation for the Green function 
in terms of intermediate states is chosen.
The saddle point integration over time variable $t^\prime$ is
intrinsic for the factorization technique, whereas
another time integration could be carried without
approximations, for instance, numerically.
The latter approach was adopted in Refs.~\cite{KOlet,KOpap}
and in Sec.~\ref{phen} of the present paper. However,
EC approximation implies saddle point integration also
over time variable $t$.
The details could be found in Appendices \ref{A1} and \ref{A2}.
Here we only indicate physical meaning of the emerging saddle points.
For $t^\prime$ variable the saddle point $t^\prime_{m \mu_0}$,
being given by Eq.~(\ref{spATI}),
is the instant of time when the electron is emitted into 
an intermediate ATI channel. The saddle point in $t$-integral, 
$t_{N \, m \mu}$, see Eq.~(\ref{sprad}), is the time of 
laser-assisted recombination 
of electron with the core, accompanied by emission of 
high-energy photon. Note that generally both  
$t^\prime_{m \mu_0}$ and $t_{N \, m \mu}$ are complex-valued.

Formula (\ref{dfin}) is important conceptually 
being the most direct and fully quantum description
of the three-step mechanism of HG. It can serve also as a practical 
computational tool. The latter statement was testified by a very good
quantitative agreement between calculations of HG rates 
by formulae (\ref{dfin}), (\ref{RN}) \cite{KOlet,KOpap} and 
the benchmark results by Becker {\it et al}\/ \cite{B} 
for HG by H$^-$ ion.
However, as discussed in Sec.~\ref{int1}, it was not explored
before what is the range of channel label $m$ that gives
substantial contribution to the rate. This  problem is addressed 
in Sec.~\ref{phen} where we show that a rather large number 
of intermediate ATI channels is to be taken into account. 
The effective summation method of Sec.~\ref{eff} transforms 
sum (\ref{sumc}) over large number of $m$-terms to the sum 
over very small number (one or two) of EC terms.
The illustrative applications of EC approach in Sec.~\ref{app}
serve to demonstrate its reliability.
EC representation is convenient also for analysis 
of general features of HG spectrum.
The latter is known to consist of three parts. The initial rapid 
decrease is followed by the {\it plateau}\/ domain and the rapid 
{\it cutoff}\/ region. Being interested in the generation of rather 
high harmonics, we do not consider
below the initial low-$N$ part of the spectrum. The upper
boundary $N_{\rm b}$ of the plateau 
is given by the known expression \cite{KulanderA,KulanderB,Lew,B}
\begin{eqnarray} \label{bound314}
N_{\rm b} \, \omega = \frac{1}{2} \kappa^2 + 3.17 \, U_p ~. 
\end{eqnarray}
In Sec.~\ref{boundary} we show how this important result follows
from the three-step mechanism cast in the EC form. 
Section \ref{conc} contains concluding discussion.

\section{Composition of HG amplitude from ATI contributions}
\label{phen}

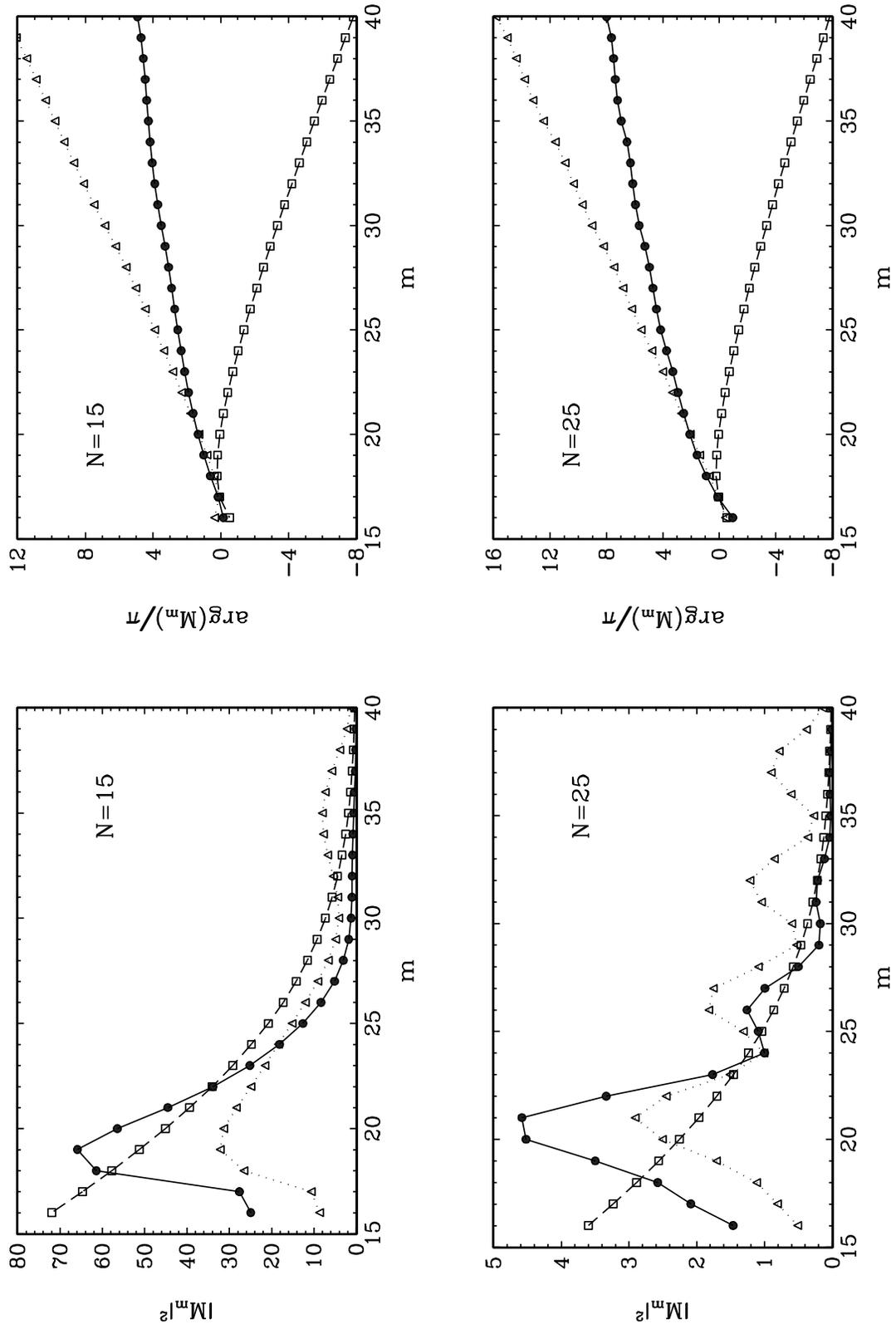
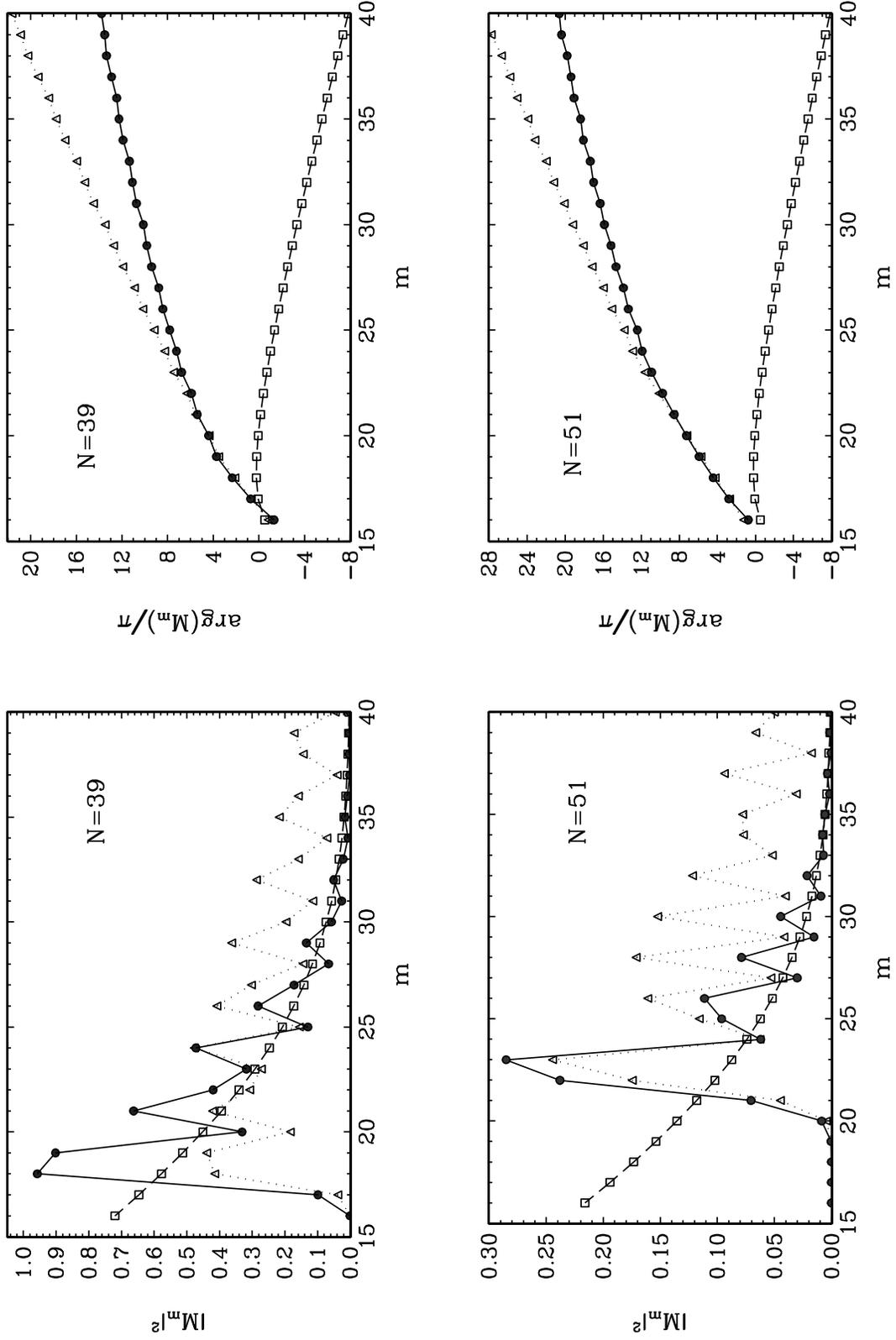
\begin{figure} 
\caption{\label{cont}
Contributions to the HG amplitude $d^+_N$
from different ATI $m$-channels according to the expression
(\protect \ref{dfin}). The H$^-$ ion in the laser field 
with the frequency $\omega = 0.0043$ and intensity 
$I$ = 10$^{11}$ W/cm$^2$
produces harmonic of the order $N$ indicated in the plots.
The square moduli $|M_m|^2$ and phases $\arg (M_m)$
of three different amplitudes $M_m$ are pictured.
The closed circles show the squared moduli and phases  
of the terms 
$d^+_{Nm} \equiv A_{m \, \mu_0} \:B_{N \, m \mu_0}$
in the sum (\protect \ref{dfin}) over ATI channel label $m$.
We present also the squared moduli and phases
of the physically meaningful factors that constitute
$d^+_{Nm}$: ATI amplitude $A_{m \, \mu_0}$ (open squares)
and PLAR amplitude $B_{N \, m \mu_0}$ (open triangles).
For graphical representation the convenient scaling factors 
are introduced in the plots of squared moduli, namely 
as $|M_m|^2$ we show everywhere $10^8 \, \left| 2 d^+_{Nm} \right|^2$, 
and also:
for $N=15$ -- $10^6 \, \left| A_{m \, \mu_0} \right|^2$ and
$10^4 \, \left| B_{N \, m \mu_0} \right|^2$;
for $N=25$ -- $5 \cdot 10^4 \, \left| A_{m \, \mu_0} \right|^2$ and
$10^4 \, \left| B_{N \, m \mu_0} \right|^2$;
for $N=39$ -- $10^4 \, \left| A_{m \, \mu_0} \right|^2$ and
$10^4 \, \left| B_{N \, m \mu_0} \right|^2$;
for $N=51$ -- $3 \cdot 10^3 \, \left| A_{m \, \mu_0} \right|^2$ and
$10^4 \, \left| B_{N \, m \mu_0} \right|^2$.
}
\end{figure}

Representation (\ref{dfin}) of HG amplitudes $d^+_N$
as a sum over ATI contributions is characterized by a number 
of interesting features. We illustrate them by Fig. \ref{cont}
where different terms in the sum (\ref{sumc}) are
quantitatively presented for HG by H$^-$ ion in 
the laser field with the frequency $\omega = 0.0043$
and the intensity $I$ = 10$^{11}$ W/cm$^2$. We show 
separately the squared moduli and phases for the 
ATI amplitude $A_{m \, \mu_0}$ (open squares),
PLAR amplitude $B_{N \, m \mu_0}$ (open triangles) 
and the resulting term in the sum (\ref{sumc}) 
$d^+_{Nm} 
%\equiv A_{m \, \mu_0} \:B_{N \, m \mu_0}
$ 
(closed circles) in dependence on $m$ for some
representative values of the harmonic order $N$.
Namely, $N= 15, \: 25$ lie in the plateau domain of HG spectrum
whereas $N= 39, \: 51$ are in the cutoff region.
Only open intermediate ATI channels are included in summation 
(\ref{sumc}) that in the present case means $m \geq 16$.

The following observations could be made.

\begin{itemize}

\item
The number of $m$-terms giving substantial contribution
to the sum (\ref{sumc}) is essentially independent of
$N$ being about 15 in the example under consideration. 

\item
The domain of $m$ giving substantial contribution
is defined, first, by the monotonous decrease of 
$\left| A_{m \, \mu_0} \right|^2$ with $m$ that ensures 
effective cut-off from the high-$m$ side.
The second factor $\left| B_{N \, m \mu_0} \right|^2$
has more complicated behavior. Indeed, it oscillates 
with $m$ and has substantial magnitude in the much broader 
range of $m$ than $\left| A_{m \, \mu_0} \right|^2$ 
(this issue is discussed in
more detail in the Appendix \ref{LAR}). In HG process
it is important that for high harmonic order $N$ 
the factor $\left| B_{N \, m \mu_0} \right|^2$ is strongly
suppressed in the low-$m$ region (i.e. near ATI 
threshold, see Figs. \ref{cont}c,d). Therefore, as $N$ 
increases, the domain of important contributions to the sum 
shifts to higher $m$.

\item
Typical values of the {\it modulus squared components}\/ 
$\left| d^+_{Nm} \right|^2$ decrease with the harmonic order 
$N$ increasing. This is explained by decrease in 
the typical values of the $\left| B_{N \, m \mu_0} \right|^2$ 
factor (note that $\left| A_{m \, \mu_0} \right|^2$ does 
not depend on $N$). Importantly, this relatively slow decrease 
cannot explain the cutoff in the spectrum of {\it modulus squared
amplitudes}\/
$\left| d^+_N \right|^2$. For instance, as $N$ varies from
39 to 51 the typical values of $\left| d^+_{N m} \right|^2$
decrease only by a factor about 3, whereas the resulting 
$\left| d^+_N \right|^2 = \left| \sum_m d^+_{N m} \right|^2$
decreases by seven orders of magnitude !
This clearly indicates that the high-$N$ cutoff of HG rate  
is due to the strong {\it cancellation of terms}\/ in the sum 
(\ref{sumc}), i.e. that the {\it coherence or phase effects}\/ 
are crucial. 

\item
The plots of the phases show that 
$\arg \left( A_{m \, \mu_0} \right)$
decreases with $m$ whereas 
$\arg \left( B_{N \, m \mu_0} \right)$ 
increases. The resulting phase
$\arg \left( d^+_{Nm} \right) =
\arg \left( A_{m \, \mu_0} \right) +
\arg \left( B_{N \, m \mu_0} \right)$
grows with $m$ rather rapidly.
For larger $N$ variation of phases becomes faster.
For instance, in the case $N=51$ the phase   
$\arg \left( d^+_{Nm} \right)$ increases by about 10$\pi$
along the substantial $m$-domain ($20 < m < 35$).
Namely the rapid phase variation governs cancellations
in the sum (\ref{sumc}) and rapid decrease of the
HG rates in the fall-off region.

\end{itemize}

The rapid variation of phases is generally characteristic 
for the semiclassical dynamics. Therefore one can suggest 
that in our problem some semiclassical-type
method of summation over ATI contributions could be
developed that reflects the physics of the process.
This program is implemented in the next section.

\section{Effective ATI Channels}
\label{eff}

\subsection{Poisson summation of ATI channel contributions}
\label{sum}

Further in the basic formula (\ref{dfin}) we employ for 
both ATI and PLAR amplitudes the saddle-point 
approximation given by formulas (\ref{A}) and (\ref{Bsprep}), 
respectively. This representation possesses important advantage 
of explicitly exposing large phases of amplitudes which are of
crucial significance in coherent summation as revealed in previous
section. The phase factors have form of exponents of classical
action for the ATI and LAR processes.
By using these expressions we present the HG amplitude 
(\ref{dfin}) as
\begin{eqnarray} \label{dfinp}
d^+_N = 2 \, \, \sum_m \: \sum_\mu  
{\cal Q}_{m \mu} \:
\exp \left[ - i {\cal S} \left(t_{m \mu}, t^\prime_{m \mu_0} 
\right) \right] ~.
\end{eqnarray}
Bearing in mind importance of phase factors we write down \
them explicitly in Eq.~(\ref{dfinp}), introducing
\begin{eqnarray} \label{genS}
{\cal S} (t, t^\prime) = S(t) - S(t^\prime) + \Omega t =
\frac{1}{2} \int^{t}_{t^\prime} 
d \tau \left[\left({\bf K}_m + 
\frac{{\bf F}}{\omega} \sin \omega \tau \right)^2
-E_a \right] + \Omega t ~,
\end{eqnarray}
whereas all the rest is collected in the pre-exponential
factor ${\cal Q}_{m \mu}$:
\begin{eqnarray} \label{Q}
{\cal Q}_{m \mu} = - \frac{\omega^4}{2 \pi} \, A_a \, \Gamma(1+\nu/2)
\, 2^{\nu/2} \, \kappa^\nu \, Y_{lm_{\rm az}}(\hat{{\bf p}}_{\mu_0}) \:
%\nonumber \times \\ \times   
\frac{1}
{F \left(\cos \omega t^\prime_{m \, \mu_0}  - 
\cos \omega t_{m \, \mu} \right) } \:
\nonumber \times \\ \times \: 
\tilde{\phi}_a^{(\epsilon)} \left( - {\bf K}_m - 
\frac{{\bf F}}{\omega} \sin \omega t_{m \mu} \right)
%\times \nonumber \\ \times
\frac{1}
{\sqrt{S^{\prime \prime}(t^\prime_{m \mu_0})^{\nu+1}
S^{\prime \prime} \left(t_{m \mu} \right)
}} ~. 
\end{eqnarray}
${\cal S}(t, t^\prime)$ has an appealing meaning of action for HG
processes expressed as a sum of actions for constituent
ATI and LAR processes. 

From the pragmatic point of view the attractive feature of 
the sum representations (\ref{dfin}) or (\ref{dfinp})
is obvious: each term in the sum has a clear and simple analytical
expression, that helps greatly in numerical calculations.
However, since the number of essential terms in the sum is
large, as discussed in Sec.~\ref{phen}, the behavior of
the sum may differ drastically from behavior of its individual terms.
This makes difficult an interpretation of the results which follow
from (\ref{dfin}) or (\ref{dfinp}).To overcome this disadvantage 
it is highly desirable to find some other representation for the
amplitude which would be devoid of extended summations. One can
obtain guideline for practical implementation of this idea by recalling
the fact that summation in (\ref{dfin}) or (\ref{dfinp}) runs
over the {\it spectrum}\/ of intermediate ATI states. From the
general physical principles we know that when the spectrum covers
a broad range of energies it could be advantageous to work within
the time-dependent picture. Mathematically this implies the Fourier
transformation which replaces the spectral quantities $m\omega$ with 
appropriate intervals of time.

Having these arguments in mind, we carry out the Poisson
summation in (\ref{dfinp}) that could be looked at as desired ``Fourier transformation''
\begin{eqnarray} \label{ps}
d^+_N = 2 \, \, \sum_{j=-\infty}^{\infty} \:
\int_{ - \infty}^\infty dm \: \sum_\mu  
{\cal Q}_{m \mu} \:
\exp \left[ - i {\cal S} \left(t_{m \mu}, t^\prime_{m \mu_0} \right) 
- 2 \pi i j m \right] ~.
\end{eqnarray}
Transition from (\ref{dfinp}) to (\ref{ps}) amounts to replacing
summation over the integer $m$ by integration over related continuous 
variable. The price is introducing an extra (and infinite) summation 
over an integer $j$. The transformation is worthy if the latter
sum effectively contains less terms than the original 
$m$-sum (\ref{sumc}), i.e., converges more rapidly.

\subsection{Saddle point method and effective channel representation}
\label{spmeth}

Our next step is evaluation of integrals over $m$ in (\ref{ps}).
As demonstrated in Sec.~\ref{phen}, the phase of the term $d^+_{Nm}$
in the sum (\ref{sumc}) varies rapidly with the summation index $m$.
Physically this follows from a multiphoton nature of the process 
under consideration. We presume that the exponent in (\ref{ps}) 
is responsible for this, whereas the pre-exponential factor 
${\cal Q}_{m \mu}$ varies slowly. This allows us to use
the saddle point method to carry out explicitly the integration 
over the ATI channel label $m$. The position of the saddle 
point(s) is governed by equation
\begin{eqnarray} \label{spm}
\frac{d}{dm} \: 
{\cal S} \left(t_{m \mu}, t^\prime_{m \mu_0}\right) 
= - 2 \pi j ~.
\end{eqnarray}
When taking derivative in Eq.~(\ref{spm}) one has to remember that
there are two sources of $m$-dependence: first, via 
$t^\prime_{m \mu_0}$ and $t_{m \mu}$ that are the integration 
limits in the definition (\ref{genS}), and, second, via
$K_m$ that enters the integrand in (\ref{genS}).
However the situation is drastically simplified by the fact that
\begin{mathletters}
\begin{eqnarray} \label{sptt}
\left. \frac{\partial}{\partial t^\prime} \, 
{\cal S}(t, t^\prime) \, \right|_{\, t^\prime = t^\prime_{m \mu_0}} = 0 ~,
\end{eqnarray}
\begin{eqnarray} \label{sptt2}
\left. \frac{\partial}{\partial t} \, 
{\cal S}(t, t^\prime) \, \right|_{t = t_{m \mu}} = 0 
\end{eqnarray}
\end{mathletters}
since $t^\prime_{m \mu_0}$ and $t_{m \mu}$ are the saddle points 
in integration respectively over $t^\prime$ and $t$ variables 
as discussed briefly in Sec.~\ref{Int} and in more detail in
Appendices \ref{A1} and \ref{A2}, see formulae (\ref{seqa})
and (\ref{spt}). Therefore Eq.~(\ref{spm}) takes a compact 
form ($ dK_m/dm = \omega/K_m$)
\begin{mathletters}
\begin{eqnarray} \label{spmm}
\left[
K_m \left(t - t^\prime + j T \right) + 
\frac{F}{\omega} \int_{t^\prime}^t \sin \omega \tau \, d \tau 
\right]_{t^\prime = t^\prime_{m \mu_0} ;
\: \: t = t_{m \mu}}
= 0  ~, 
\end{eqnarray}
or, more explicitly,
\begin{eqnarray} \label{spmmon}
K_m \left(t_{m \mu}- t^\prime_{m \mu_0} + j T \right) =
\frac{F}{\omega^2} 
\left( \cos \omega t_{m \mu} - \cos \omega t^\prime_{m \mu_0}\right) ~.
\end{eqnarray}
\end{mathletters}
It is to be considered together with formulae (\ref{spATI})
and (\ref{sprad}) defining positions of saddle points in 
integration over $t^\prime$ and $t$ variables respectively. 
%Here we rewrite these equations as
%\begin{eqnarray} \label{sptprime}
%\sin \omega t^\prime_{m \mu_0} + K_m \, \frac{\omega}{F}
%= i \kappa \, \frac{\omega}{F} ~,
%\\ \label{sptexpl}
%\sin \omega t_{m \mu} + K_m \, \frac{\omega}{F} = 
%\sqrt{2N\omega - \kappa^2} \: \frac{\omega}{F} ~.
%\end{eqnarray}
The unknown variable to be defined is the ATI label $m_c(N,j)$. 
As shown below, only complex-valued solutions are possible.
Note that $m$ enters (\ref{spmmon}), (\ref{spATI}) and (\ref{sprad}) 
only via $K_m$ (\ref{Kmnm}).
Therefore solution of the saddle point equation (\ref{spmmon}) 
amounts to finding complex-valued translational momentum 
$K_{m_c}(N, j)$ of electron in the intermediate ATI channel
that gives major contribution to the generation of $N$-th harmonics. 

Within the saddle point approximation formula (\ref{ps}) is 
reduced to
\begin{eqnarray} \label{pssad}
d^+_N = 2 \, \, \sum_{j=-\infty}^{\infty} \: 
\sum_{m_c} \:
\sum_\mu {\cal Q}_{m_c \mu} \: 
\sqrt{\frac{2 \pi}{ i {\cal S}^{\prime \prime}_{m_c}}} \: 
\exp \left[ - i {\cal S} 
\left( t_{m_c \mu}, t^\prime_{m_c \mu_0} \right) 
- 2 \pi i j m_c \right] ~, 
\end{eqnarray}
where
\begin{eqnarray}
{\cal S}^{\prime \prime}_{m_c} \equiv
\left. \frac{d^2}{d m^2} {\cal S}
\left(t_{m \mu}, t^\prime_{m \mu_0} \right) \right|_{m = m_c}
= \frac{\omega^2}{K_m^2} 
\left( t_{m_c \mu} - t^\prime_{m_c \mu_0} + jT \right) 
+ \nonumber \\ +
\frac{\omega}{K_{m_c}}
\left( \sqrt{2N\omega-\kappa^2} \: \frac{d t_{m \mu}}{dm}
- i \kappa \, \frac{d t^\prime_{m \mu_0}}{dm} \right)_{m=m_c} ~.
\end{eqnarray}
The necessary derivatives 
%$d t_{m \mu}/dm$ and $dt^\prime_{m \mu_0}/dm$ 
are straightforwardly derived using formulas (\ref{spATI}) 
and (\ref{sprad}), respectively, as
\begin{eqnarray}
\frac{d t^\prime_{m \mu_0}}{dm} =  - \, \frac{\omega}
{K_m F \cos \omega t^\prime_{m_c \mu_0}} ~,
\quad \quad \quad
\frac{d t_{m \mu}}{dm} = - \, \frac{\omega}
{K_m F \cos \omega t_{m_c \mu}} ~.
\end{eqnarray}

Even more appealing form of expression (\ref{pssad}) is obtained
if one reexpresses the constituent factors in terms of ATI and 
LAR amplitudes:
\begin{eqnarray} \label{pssad1}
d^+_N & = & 2 \, \, \sum_{j} \: 
\sum_{m_c} \: 
A_{m_c \, \mu_0}({\bf K}_{m_c}) \,
%\, B_{N \, m_c \mu_0}({\bf K}_{m_c}) ~,
\frac{1}{R_{m_c \, \mu \, \mu_0}} \, C_{N m_c}({\bf K}_{m_c}) \,
\Xi(m_c, j) ,
\\
\Xi(m_c, j) & \equiv &
\sqrt{\frac{2 \pi}{ i {\cal S}^{\prime \prime}_{m_c}}} \,
\exp( - 2 \pi i j m_c) ~.
\end{eqnarray}
Here we omit summation over $\mu$ implying that it is absorbed 
into the sum over $m_c$, since each saddle point
$m_c$ is obtained for some choice of label $\mu$.
The form of expansion (propagation) factor 
\begin{mathletters}
\begin{eqnarray} \label{prop}
\frac{1}{R_{m \, \mu \, \mu_0}} = \frac{\omega^2}{F
\left( \cos \omega t_{m \mu} - \cos \omega t^\prime_{m \mu_0} \right)} 
\end{eqnarray} 
is a natural specification of the more general expression
obtained earlier \cite{KOpap,KOlet} and cited in Eq.~(\ref{prop1}).
In particular, at the saddle point on can use Eq.~(\ref{spmmon}) 
to rewrite propagation factor  as 
\begin{eqnarray} \label{prop2}
\frac{1}{R_{m_c \, \mu \, \mu_0}} =
\frac{1}{K_{m_c} \left(t_{m_c \mu}- t^\prime_{m_c \mu_0} + j T \right)} ~.
\end{eqnarray} 
Formula (\ref{pssad1}) provides the most concentrated 
expression of the result of present work. Apart from 
the smooth factor $\Xi(m_c, j)$, it is fully analogous 
to our starting point expression (\ref{dfin}), but with 
summation over large number of intermediate ATI channels $m$
replaced in Eq.~(\ref{pssad1}) by summation over small number 
of effective channels $m_c$. Note that the factorization 
of propagation and LAR amplitudes, as given by formula 
(\ref{prfact}), appears as an additional approximation
in the framework of our previous approach \cite{KOlet,KOpap}
where the $t$-integration was carried out numerically. 
In the EC formulation the factorization is a necessary feature, 
due to the saddle-point method applied to calculation of integral 
over time variable $t$, see Appendix \ref{A2}.
The form (\ref{prop}) of the expansion factor is more accurate 
than the approximation
\begin{eqnarray} \label{propap}
\frac{1}{R_{m \, \mu_0}} = - \, \frac{\omega^2}{F
\cos \omega t^\prime_{m \mu_0} } 
\end{eqnarray} 
\end{mathletters}
discussed (within unessential sign) earlier \cite{KOlet,KOpap}
and implied in formula (\ref{prfact}) of the present paper.

The transparent interpretation of the saddle point equation 
(\ref{spmm}) is based on the fact that 
\begin{eqnarray} \label{disp}
z \left(t, t^\prime ; \, p \right) = p
\left(t - t^\prime + j T \right) + 
\frac{F}{\omega} \int_{t^\prime}^t \sin \omega \tau \, d \tau 
\end{eqnarray}
is the electron displacement along $z$-axis in course of 
its wiggling motion in the laser field as time varies from 
$t^\prime$ to $t$. The axis is directed along the electric
field vector {\bf F}; the electron translational momentum
${\bf p}$ has the same direction.
% and its magnitude corresponds to the $m$-th ATI channel.
In terms of $z \left(t, t^\prime ; \, p \right)$ (\ref{disp}) 
the saddle point equation (\ref{spmm}) can be equivalently 
rewritten as
\begin{eqnarray} \label{z0}
z \left(t_{m \mu} + j T , \: t^\prime_{m \mu_0} ; \, K_m \right) = 0 ~.
\end{eqnarray}
This equation has a lucid physical meaning.
In the semiclassical picture of HG, as implemented by the present 
theory,
%within the saddle point method, 
the electron emerges
from the under barrier at the instant of time $t^\prime_{m \mu_0}$
into the $m$-th channel of ATI continuum 
and undergoes backward transition into the bound state
with emission of the $\Omega$ photon at time
$t_{m \mu}$. Both these events occur relatively close
to the atomic core. Eq.~(\ref{z0}) is the condition that 
after propagation in the laser field the electron returns 
{\it exactly}\/ \cite{ftnt1} to the emission point. 
It accounts to the fact that the propagation time 
could be augmented by additional $j$ laser field periods $T$. 
The integer parameter $j$ conjugate to $m$ in the 
``Fourier transformation'' (\ref{ps}) can be named a 
{\it recursion number}. As discussed above, in fact 
equation (\ref{z0}) defines an appropriate electron momentum 
$K_{m_c}$. The latter proves to be complex-valued as shown below.

The return condition (\ref{z0}) is universal in the sense
that it does not contain explicitly the harmonic 
order $N$ (the latter parameter implicitly defines 
the return time $t_{m \mu}$).
The condition of return is intrinsic in the physical picture
of the three-step process. Being constituent part of the
atomic antenna idea of Ref.~\cite{Ku87}, it appeared in some 
form in previous studies by other authors. The most close links
could be established with Ref.~\cite{Lewphase}.
This paper employs a different representation that does
not appeal to the intermediate ATI channels. Nevertheless 
the saddle point method applied in these calculations also 
leads to the equation for the saddle points with the same
physical meaning. Moreover, the same number of saddle points 
plays substantial role (namely, one point at the cutoff region
and two points at the plateau domain, see below). Unfortunately the 
saddle point evolution with varying harmonic order $N$ was 
not described in detail in the cited paper. Note also that the
cited paper provides the HG rates only within an adjustable
normalization factor.

As already mentioned, the summation over $j$ 
in the formulas (\ref{ps}), (\ref{pssad}) or (\ref{pssad1}) 
reflects the fact that the time interval between the 
emission of electron and emission of the high-energy
photon could be different; namely, several ($j$) periods 
of the laser field could be added to it. All these events 
add up coherently as show Eqs.~(\ref{pssad}) or (\ref{pssad1}). 
The shortest time 
interval corresponds to $j=0$. Physically one can anticipate
that the related contribution prevails because of a spread
of electron wave packet. Mathematically the term with $j=0$ 
is obtained by plain replacement of the summation over 
$m$ in Eq.~(\ref{dfin}) by integration. Further on we concentrate
on calculation of this term and consider at first 
(Sec.~\ref{boundary}) the related saddle points 
$m_c(N) \equiv m_c(N, j=0)$.

Let us discuss now a restriction on the summation index $j$ in 
(\ref{pssad}). In the saddle point method one deforms the 
integration contour so that it passes via the saddle points;
only close vicinities of the saddle point effectively
govern the integral magnitude. Generally, the shifted 
contour passes only via some subset of the saddle points 
available; only the points from this subset contribute 
to the integral.
We deal with multi-dimensional problem (integration over 
variables $m, \, t, \, t^\prime$) that casts analysis 
of the integration contour deformation as a complicated task.
To circumvent this obstacle we adopt  an approach based on 
simple physical arguments. As discussed above,
the recursion number $j$ counts a number of periods 
of time that elapse between
the electron escape and its return to the atom. 
The three-step mechanism presumes that the moment of escape must
precede the moment of return. This condition 
can only be satisfied for non-negative values of $j$. 
Basing on this  physical argument we  assume that
the saddle points $m_c$ contribute to the integral 
only for non-negative $j\ge 0$.
%as is indicated in summation over $j$ in (\ref{pssad}).

Saddle points $m_c$ which contribute to summation over $m_c$ 
in formula (\ref{pssad}) should satisfy additional condition 
\begin{equation} \label{im<0}
{\rm Im} \left[ {\cal S} 
\left( t_{m_c \mu}, t^\prime_{m_c \mu_0} \right) 
+ 2 \pi  j m_c \right] < 0~,
\end{equation}
which ensures that the factor in the exponent in 
the right-hand side of Eq.~(\ref{pssad}) always reduces 
the absolute value of the amplitude. Again we do not attempt 
to validate (\ref{im<0}) mathematically, but note instead 
that this condition is very similar to a conventional 
restriction on the resonances in stationary processes.
Remember that the energy $E_0$ of a conventional resonance has 
the real and imaginary parts. The later one describes the width
of the resonance ${\rm Im} E_0 = -\Gamma/2<0 $, and must be negative. 
In the time-dependent formalism the resonance 
contribution to the amplitude is given by the factor 
$\exp (- i E_0 t)$, where $t>0$ is the time elapsed since 
the resonance intermediate state has been excited. 
The condition ${\rm Im} E_0 <0$ selects one of
two complex-conjugate poles of the $S$-matrix, or Green function,
thus ensuring the decay of the resonance, i.e., that 
$\left| \exp (- i E_0 t) \right| < 0$.
Similar physical meaning has Eq.~(\ref{im<0}).

Eqs.~(\ref{pssad}) or (\ref{pssad1}) are the major result of this paper.
They implement the objective formulated in Sec.~\ref{sum} to carry out 
the ``Fourier transformation'' of Eq.~(\ref{ps}). 
The latter expression includes summation over the physical, 
discrete spectrum $m$ of the intermediate ATI states.
In contrast, Eq.~(\ref{pssad}) refers to some complex-valued $m_c$
which label EC. Additionally, Eq.~(\ref{pssad}) includes summation 
over recursion number $j$, that is a number of laser periods $T$
that elapse between the ionization and HG generation. It is important 
that the sum over $j$ converges well. The reason for this
originates from the discussed above fact of the slow convergence
of the spectral representation (\ref{ps}). It is a general, 
well known fact that if the spectral pattern is broad, then 
the ``Fourier transformation'' should be well localized. 
Same argument can be presented from another point of view.
The shortest (positive) time interval between the ionization and
HG emission corresponds to $j=0$. One can anticipate
that the related contribution prevails because the spread
of the electron wave packet should significantly diminish 
contributions of events with larger $j$.
Presuming that this argument is correct, we will replace 
in the applications (Sec.~\ref{app}) summation
over $j$ in Eq.~(\ref{pssad1}) by the only term
with $j=0$. Good numerical results obtained
illustrate the fact that the term $j=0$ really dominates.
For $j=0$ we find only few (one or two) operative EC $m_c$.
Thus the sum (\ref{pssad1}) over EC comprises very few terms
that is convenient for interpretation of the results.

The physical background of the mathematical transformations 
above can be briefly summarized as follows. The electron 
motion in a laser field in the vicinity of an atom satisfies 
the adiabatic condition because the number of laser quanta 
absorbed and emitted in the HG process is large.
The HG amplitude contains a large phase that is identical to the
classical action ${\cal S}(t,t^\prime)$ (\ref{genS}).
This phase varies rapidly with all the parameters that
govern the electron  propagation in the intermediate state.
Therefore the major contribution to the event comes 
from such situations in which the phase is stationary,
i.e. from the saddle points. The phase depends on the initial 
moment of virtual ionization $t^\prime$, the final moment of 
the harmonic emission $t$ and the energy of the electron
in the intermediate state. Eqs.~(\ref{sptt}), (\ref{sptt2}), 
(\ref{spm}) represent the saddle-point conditions over these 
three variables. Altogether they define the two moments of 
time and the electron energy in the intermediate state. 

%This discussion  shows that the electron propagation
%in the intermediate state
%can be conveniently described by some complex-valued energy.
%We can consider a state with this energy as a particular,
%specific for the multiphoton problem resonance, calling it
%Resonance in the ATI spectrum, RATI for short.
%Note that all properties of this resonance depend entirely on the laser field,
%being independent on the details of an atomic potential. 
%It should be  mentioned also
%that RATI, as discussed above,  is  specially 
%designed to represent a set of the ATI states 
%which arise in the intermediate
%state of HG. We will discuss later on that
%the idea of RATI  can also be applied to other three-step
%process, always describing the intermediate
%step of the process. 
%It is not applicable for the description of
%ATI levels in the final state.

\subsection{Analysis of saddle point equation}
\label{boundary}

By summing and subtracting Eqs.(\ref{spATI}) and (\ref{sprad}) 
we obtain
\begin{mathletters}
\begin{eqnarray} \label{sptsum}
\sin \omega t_{m \mu} + \sin \omega t^\prime_{m \mu_0} +  
2 K_m \, \frac{\omega}{F} & = & 2 {\cal Z} ~, 
\\ \label{sptdif}
\sin \omega t_{m \mu} - \sin \omega t^\prime_{m \mu_0} 
& = & 2 {\cal Z}^\prime ~,
\end{eqnarray}
\end{mathletters}
with 
\begin{eqnarray}
{\cal Z} \equiv \frac{1}{2} \left( 
\sqrt{2N\omega - \kappa^2} + i \kappa \right) \frac{\omega}{F} ~,
\quad \quad \quad
{\cal Z}^\prime \equiv \frac{1}{2} \left( 
\sqrt{2N\omega - \kappa^2} - i \kappa \right) \frac{\omega}{F} ~.
\end{eqnarray}
If $2 N \omega > \kappa^2$, then one has ${\cal Z}^* = {\cal Z}^\prime$.
It is convenient to switch from $t^\prime_{m \mu_0}$ and $t_{m \mu}$ to
\begin{eqnarray} \label{xy}
x = \frac{1}{2} \, \omega \left( t_{m \mu} + t^\prime_{m \mu_0} \right) ~,
\quad \quad \quad
y = \frac{1}{2} \, \omega \left( t_{m \mu} - t^\prime_{m \mu_0} \right) 
\end{eqnarray}
and present (\ref{sptsum}) and (\ref{sptdif}) in the compact form 
\begin{mathletters}
\begin{eqnarray}
\sin x \, \cos y + K_m \, \frac{\omega}{F} = {\cal Z} ~,
\\
\cos x \, \sin y = {\cal Z}^\prime ~.
\end{eqnarray}
\end{mathletters}
In the same notation Eq.~(\ref{spmmon}) reads 
\begin{eqnarray}
K_m \, \frac{\omega}{F} = - \frac{ \sin x \, \sin y}{y+j \pi} ~. 
\end{eqnarray}
By excluding $K_m$ we obtain a set of two equations for two
variables $x$ and $y$:
\begin{mathletters}
\begin{eqnarray}
\sin x \left( \cos y - \frac{\sin y}{y + j \pi} \right) = {\cal Z} ~,
\\
\cos x \, \sin y = {\cal Z}^\prime ~.
\end{eqnarray}
\end{mathletters}
The next step is to exclude $x$ that after some algebra gives
a compact transcendental equation for a single variable 
$\tilde{y} = y + j \pi$:
\begin{eqnarray} \label{yex}
\left( \cot \tilde{y} - \frac{1}{ \tilde{y}} \right)^2 
\left[ \sin^2 \tilde{y} - \left( {\cal Z}^\prime \right)^2 \right] 
= {\cal Z}^2 ~.
\end{eqnarray}
As soon as $\tilde{y}$ is found, one obtains $K_m$ from
\begin{eqnarray}
\frac{\omega}{F} \, K_m =  
\frac{{\cal Z}}{1 - \tilde{y} \cot \tilde{y}} ~.
\end{eqnarray}
The variable $y$ (\ref{xy}) has a very lucid meaning:
$2y/\omega$ is the active electron excursion time in the continuum,
i.e., the time interval between the first step in the
three-step process, ATI and the last step, LAR. 
Change in the recursion number $j$ is equivalent to
change of the excursion time
$\left( t_{m \mu} - t^\prime_{m \mu_0} \right)$ by an integer
multiple of laser period $T$.
%We see that to consider various values of the recursion 
%number $j$ is the same as to choose different intervals 
%of length $\pi$ on the $\tilde{y}$-axis.
To simplify subsequent analysis we restrict it to the case $j=0$,
or $0 \leq \tilde{y} = y < \pi$, 
that gives the major contribution to the amplitude, as
discussed in Sec.~\ref{spmeth}. 

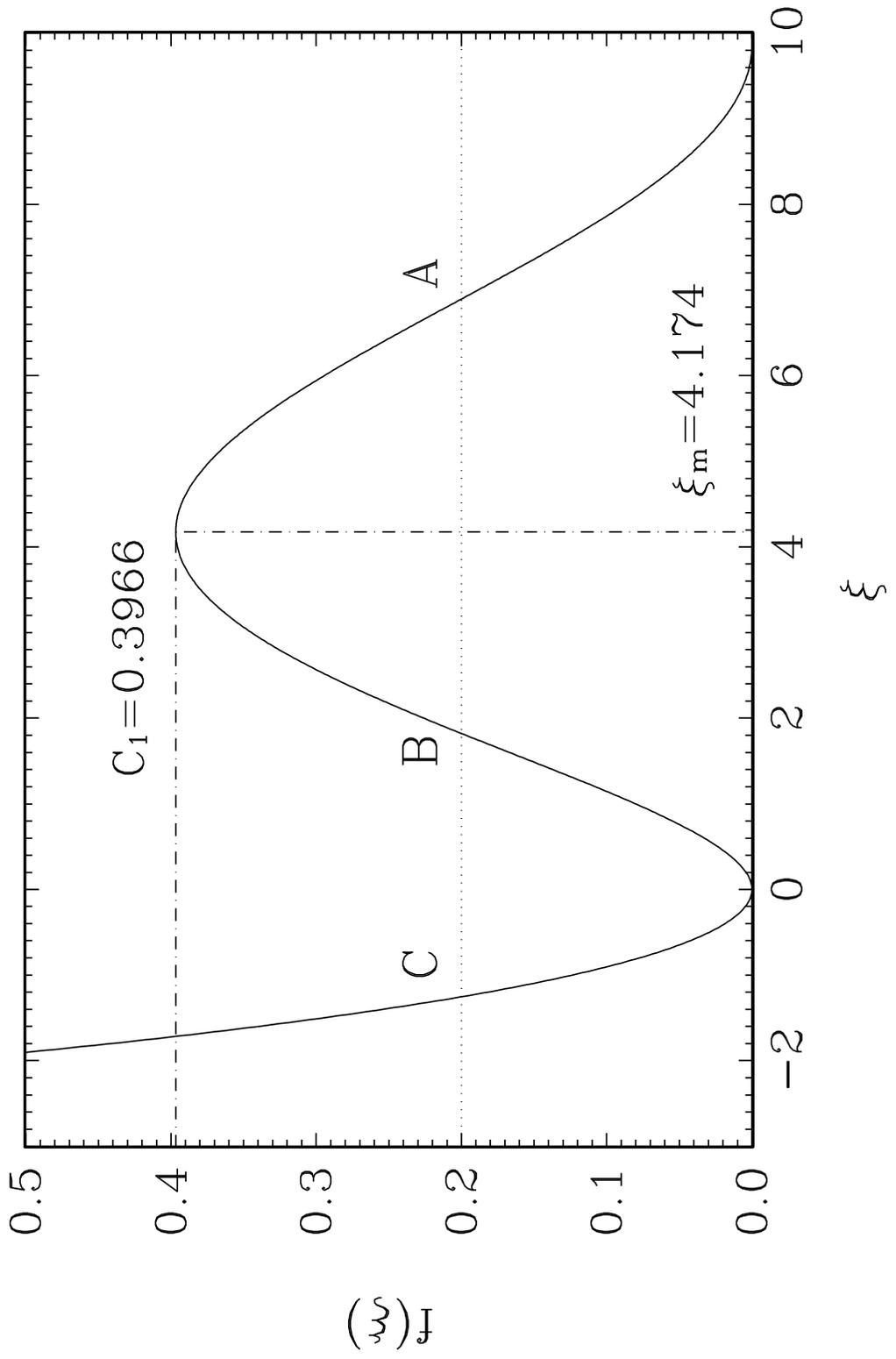
\begin{figure}
\caption{ \label{f(y)}
The universal function $f(\xi)$ (\protect \ref{f(y)def}).
}
\end{figure}

When generation of high harmonics is considered, 
$2 N \omega \gg \kappa^2$, the estimate
\begin{eqnarray} \label{est}
\sqrt{2 N \omega - \kappa^2} \gg \kappa
\end{eqnarray}
is valid that implies that one can put  
${\cal Z}^\prime \approx {\cal Z} \approx 
\sqrt{2 N \omega - \kappa^2} \, \omega / (2 F)$.
This allows us to rewrite equation (\ref{yex}) governing
saddle point positions in an approximate form
\begin{eqnarray} \label{appeq}
f(y^2) = \frac{\omega^2}{4 F^2} \left( 2 N \omega - \kappa^2 \right) ~,
%\quad \quad \quad 
\end{eqnarray}
with the universal function $f(y^2)$ of a dimensionless parameter $y$
defined as 
\begin{eqnarray} \label{f(y)def}
f(y^2) \equiv
\frac{ \left(y \cot y - 1 \right)^2 \, \sin^2 y}
{ \left(y \cot y - 1 \right)^2 + y^2 } ~.
\end{eqnarray}
A plot of this function (Fig.~\ref{f(y)}) shows
that in the interval of interest, $0 \leq \xi = y^2 < \pi^2$
it has a single maximum located at $\xi_m = 4.173$ ($y_m = 2.043$).
For sufficiently small $N$, when the right hand side of 
Eq.~(\ref{appeq}) is less than $C_1 \equiv f(\xi_m) = 0.3966$,
this equation has two solutions designated as A and B in
Fig.~\ref{f(y)} where this situation is explicitly displayed.
One can verify that these solutions satisfy condition (\ref{im<0})
and therefore both of them contribute to the HG amplitude.
Note that both solutions correspond to real but different excursion 
times $2 y /\omega$.

When $N$ increases, the roots A and B come closer and 
eventually merge at some critical value of 
$N=N_{\rm b}$ defined from
\begin{eqnarray}
\frac{\omega^2}{4 F^2} 
\left( 2 N_{\rm b} \, \omega - \kappa^2 \right) = C_1 ~,
%\quad \quad \quad
%C_1 \equiv f(y^2_m) = 0.3966 ~,
\end{eqnarray}
that is 
\begin{eqnarray}\label{boun}
N_{\rm b} \, \omega = \frac{1}{2} \kappa^2 
+ C_1 \, \frac{2 F^2}{\omega^2}
= \frac{1}{2} \kappa^2 + 3.1731 U_p ~. 
\end{eqnarray}
For larger $N>N_{\rm b}$ the solutions A and B spit again,
but this time they result in complex-valued $y$. 
The way to see this, is to note
that  in the vicinity of the maximum the function 
$f(y^2)$ behaves as $f(y^2)-f(y^2_m) \sim  - (y-y_m)^2$.
Further straightforward analyses indicates that only one
of these two complex-valued solutions satisfies condition 
(\ref{im<0}) and contributes to the amplitude, while another 
one gives no contribution. 
%The operative solution as well as the other one correspond to 
%complex-valued excursion time with large imaginary part. 

Thus the number and the character of relevant solutions 
of Eq.~(\ref{appeq}) differs for $N< N_{\rm b}$ and $N> N_{\rm b}$. 
The excursion time switches from real to complex-valued
that indicates that the HG process changes it nature from
the classically allowed to classically forbidden one. Accordingly,
as detailed in Sec.~\ref{app}, at the point $N< N_{\rm b}$
the HG spectrum undergoes a transition from the plateau to 
the cutoff region. Thus Eq.~(\ref{boun}) rederives the 
well-known upper border of the plateau domain in the HG 
spectrum (\ref{bound314}) with a slightly different coefficient 
in front of $U_p$. The validity condition for this derivation 
(\ref{est}) essentially means that the photon energy 
$N_{\rm b} \, \omega$ is much higher than the initial 
electron binding energy $\frac{1}{2} \kappa^2$. 

Before concluding this subsection we remark that one more 
solution of Eq.~(\ref{appeq}) is shown in Fig.~\ref{f(y)}, 
being labeled as C. It corresponds to negative $y^2$ and has 
simple $N$-dependence since the function $f(y^2)$ is monotonous
on the semiaxis $y^2<0$. The fact that the solution C results 
in imaginary excursion time $2 y/\omega$ indicates that it gives 
small contribution to the HG amplitude.
%and may be omitted in the subsequent analyses.

%Within the saddle point method it is important to trace
%which saddle points are operative in the contour integration.
%Only these particular saddle points are to be included into
%the summation over $m_c$ in the formula (\ref{pssad}).
%In order to understand the situation better it is useful to 
%consider the evolution of the saddle points $m_c$ in the plane
%of complex ATI label $m$ as the physical parameters vary. 

\section{ILLUSTRATIVE APPLICATIONS}
\label{app}

\subsection{Scanning harmonic order for fixed laser intensity}
\label{tra}

There are two natural and complementary outlooks on the results 
for HG process. One can fix the laser intensity $I$ and scan
the rates for harmonics of different order, as done in this
subsection, or one can fix the harmonic order $N$ and vary the 
laser intensity (Sec.~\ref{ind}). In both cases we consider HG 
by H$^-$ ion in the laser field with the frequency $\omega=0.0043$.

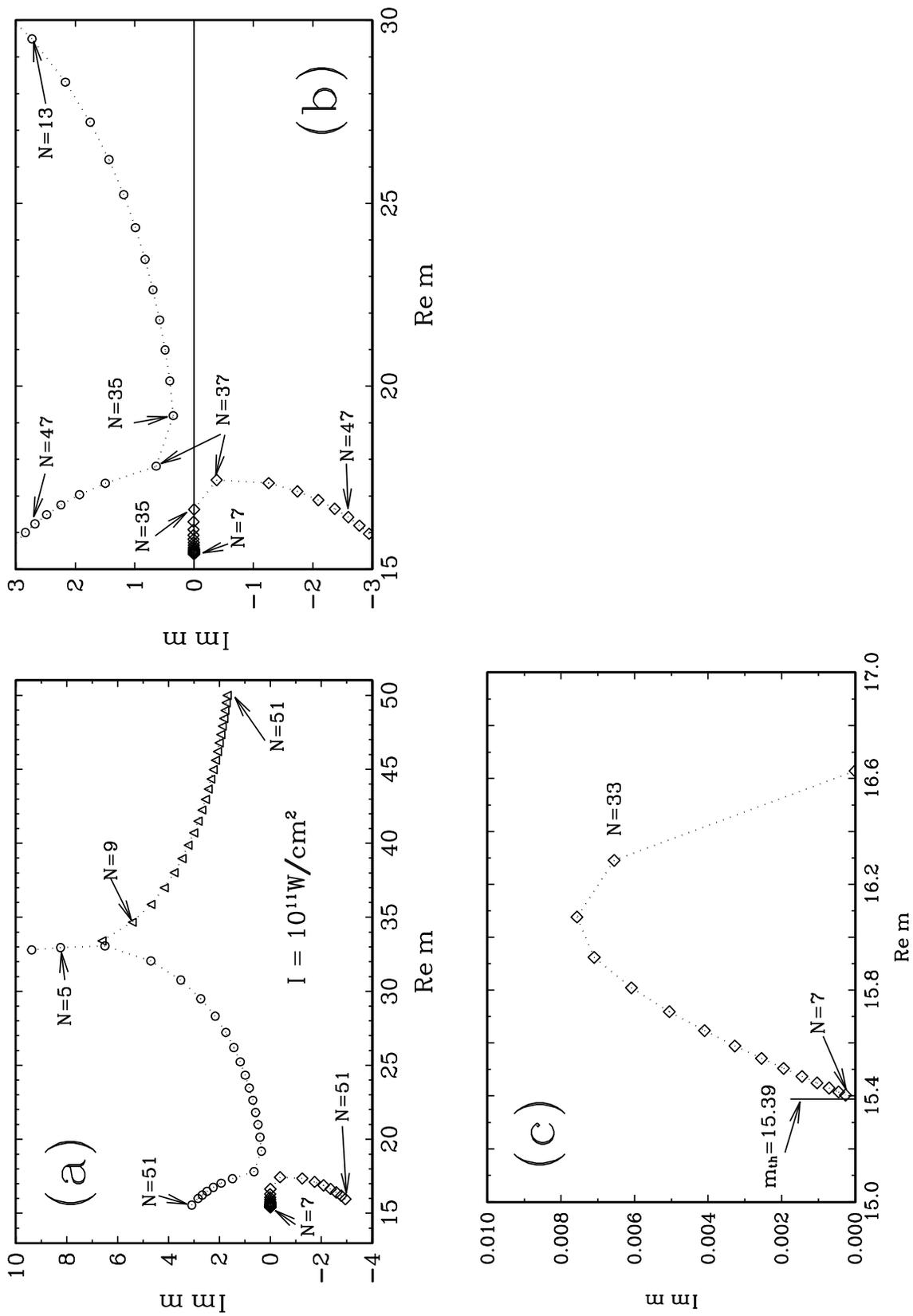
\begin{figure}
\caption{ \label{comp}
Trajectories of the saddle points $m_c(N)$ in the complex-$m$ plane 
for fixed laser intensity $I$ = 10$^{11}$ W/cm$^2$ and varying 
harmonic order $N$. The results are for H$^-$ ion 
in the laser field with the frequency $\omega = 0.0043$.
Positions of three saddle points
for odd integer $N$ are denoted respectively by diamonds, 
circles and triangles. The plot (a) gives general overview,
the plot (b) shows the region where two saddle points ``collide'',
whereas the plot (c) details behavior of the saddle point 
trajectory in the vicinity of $m = m_{\rm th}$. 
}
\end{figure}

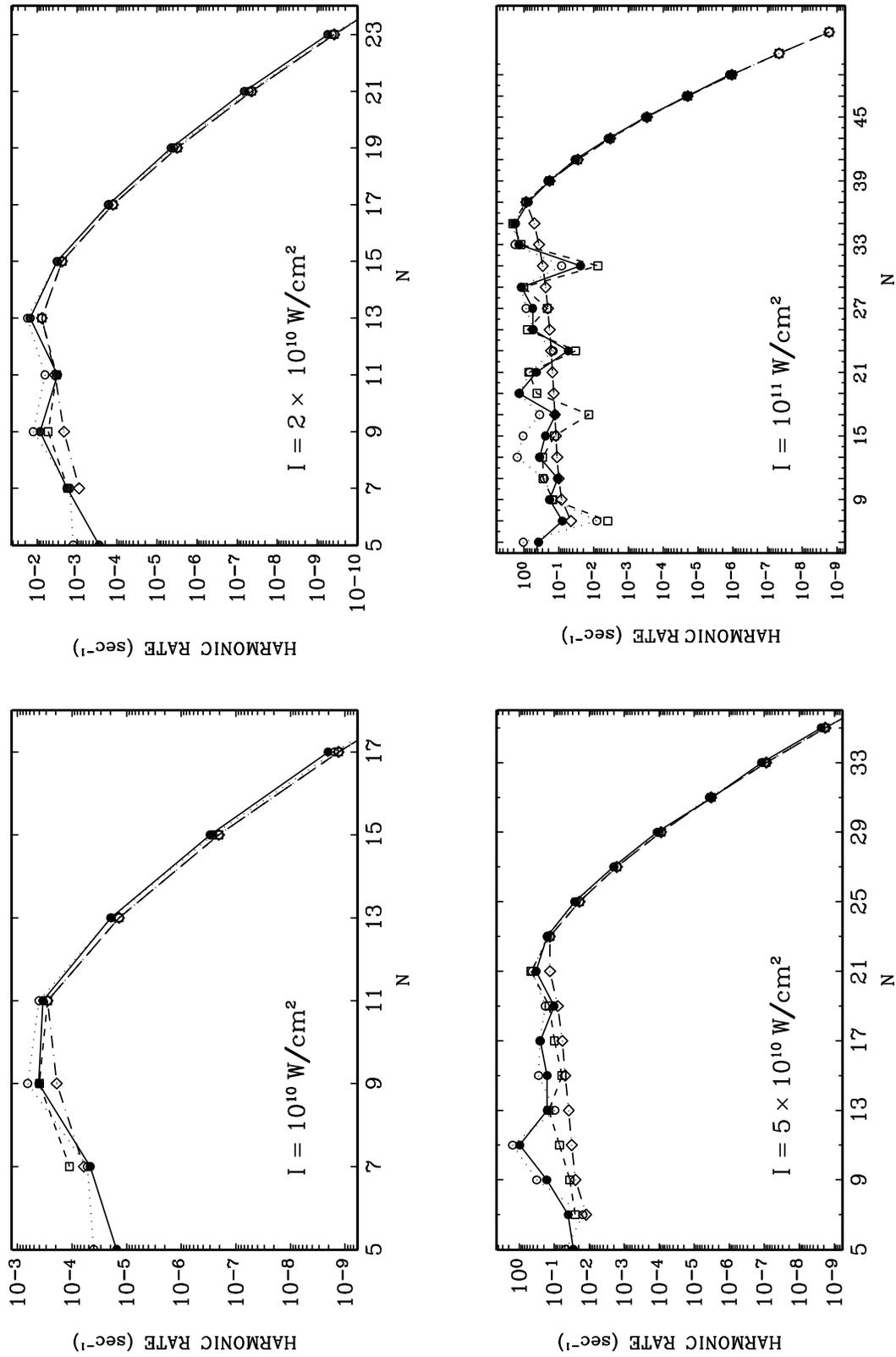
\begin{figure}
\caption{ \label{rates}
Harmonic generation rates (\protect \ref{RN})
(in sec$^{-1}$) for H$^-$ ion 
in the laser field with the frequency $\omega = 0.0043$ 
and various values of intensity $I$ as indicated in the plots.
Closed circles - results obtained by Becker {\it et al}\/
\protect\cite{B}, open circles - our calculations \protect\cite{KOpap}
in the dipole-length gauge 
%using the expression for $B_{N m \, \mu}$ and 
performing numerical summation (\protect\ref{dfin}) over 
contributions of different ATI channels, open diamonds -- 
present results within the saddle point 
approximation (\protect \ref{pssad}) for the summation over 
ATI channels contributions with a single saddle point $m_c(N)$
taken into account (namely, the saddle point shown by 
diamonds in Fig.~\protect\ref{comp}); 
open squares -- same but taking into account two saddle
points $m_c(N)$ (namely, these shown by 
diamonds and circles in Fig.~\protect\ref{comp}).
}
\end{figure}

Fig.~\ref{comp} depicts a typical pattern of evolution of 
$m_c(N)$ in the complex-$m$ plane for fixed laser field 
intensity $I = 10^{11}$ W/cm$^2$. Positions of three 
complex-valued roots of the exact saddle-point 
equation (\ref{spmmon}) are shown by symbols of different shape.
The roots move as the harmonic order $N$ varies;
we show their positions for odd (physical) values of $N$. 
For large $N$ only one root has negative imaginary part
(it is shown by diamonds in Fig.~\ref{comp})
that ensures harmonic rate decrease with increasing $N$.
Fig.~\ref{rates} shows results of the rates calculation 
in the saddle point approximation (\ref{pssad}) where only 
this single root is taken into account in the summation over $m_c$.
The cutoff region in the rates spectrum is nicely
reproduced, as well as the overall pattern in the plateau 
domain. However the single-saddle-point approximation does 
not reproduce some structures in the $N$-dependence of HG rates. 
In the plateau domain the saddle point $m_c(N)$ moves 
close to the real $m$-axis. For small $N$ this saddle point
approaches the point $m = m_{\rm th}$ on the real axis in 
the complex-$m$ plane that corresponds to the ATI threshold
\begin{eqnarray} \label{mthrdef}
m_{\rm th} = \frac{\kappa^2}{2 \omega} + \frac{U_p}{\omega} ~.
\end{eqnarray}

Transition to the cutoff region with $N$ increasing
corresponds to steep bending of the saddle point trajectory
after which it moves almost perpendicular to the real axis.
The reason of the trajectory bending is seen to be 
a ``collision'' with another saddle point shown
by the circles in Fig.~\ref{comp}.
%[see Eq.~(\ref{mthr})].
%$\left[m_{\rm th} \equiv \kappa^2/(2 \omega) + U_p/\omega \right]$.
%This point is a branch point for the analytical function
%$m_c(N)$. The saddle point trajectory moves under the 
%cut for small $N$ (see also discussion in Sec.~\ref{ind}). 
%We did not explore this domain in
%more detail since the small-$N$ part of rates spectrum is beyond 
%the scope of the present study.
In the ``collision'' region the two saddle points 
(``diamonds'' and ``circles'') abruptly change 
direction of motion forming the characteristic cross-like pattern.
The ``collision'' occurs at $N = N_{\rm col} \approx 35$.
Approximately one can identify $N_{\rm col}$ with $N_{\rm b}$.
For $N < N_{\rm col}$ the two saddle points $m_c$ are 
to be taken into account in formula (\ref{pssad}). 
As shown in Fig.~\ref{rates}, this improves the results for 
the rates in the plateau domain by producing appropriate structures 
in the $N$-dependence. Remarkably, such a two-saddle-point
calculation gives correct positions of minima and maxima
in the rate $N$-dependence, albeit the magnitudes of
the rate variation is reproduced somewhat worse; for
instance the depth of the minimum at $N=17$ is quite
strongly overestimated. Tentatively one can attribute
this to the fact that if the principal term in the
approximation for $d^+_N$ considered here proves to be
anomalously small for some $N$, then the correction terms
omitted in our calculations become relatively important.
The good overall description of the structure unambiguously
identifies its nature as a result of an interference between
the contributions coming from two effective ATI channels.

The plain saddle point approximation 
(\ref{pssad}) presumes that the saddle points are well 
separated from each other; otherwise the more complicated 
uniform approximations are to be constructed. 
In our problem this refers to the ``collision'' region
$N \approx N_{\rm col}$ where our simple approximation 
somewhat overestimate rates; however we do not resort
here to the more sophisticated mathematical constructions.
Note that the third saddle point shown by triangles in
Fig.~\ref{comp} is substantial only in calculations
for the lowest harmonics. Therefore it is not taken into
account in the present calculations.

\subsection{Scanning laser intensity for fixed harmonic order}
\label{ind}

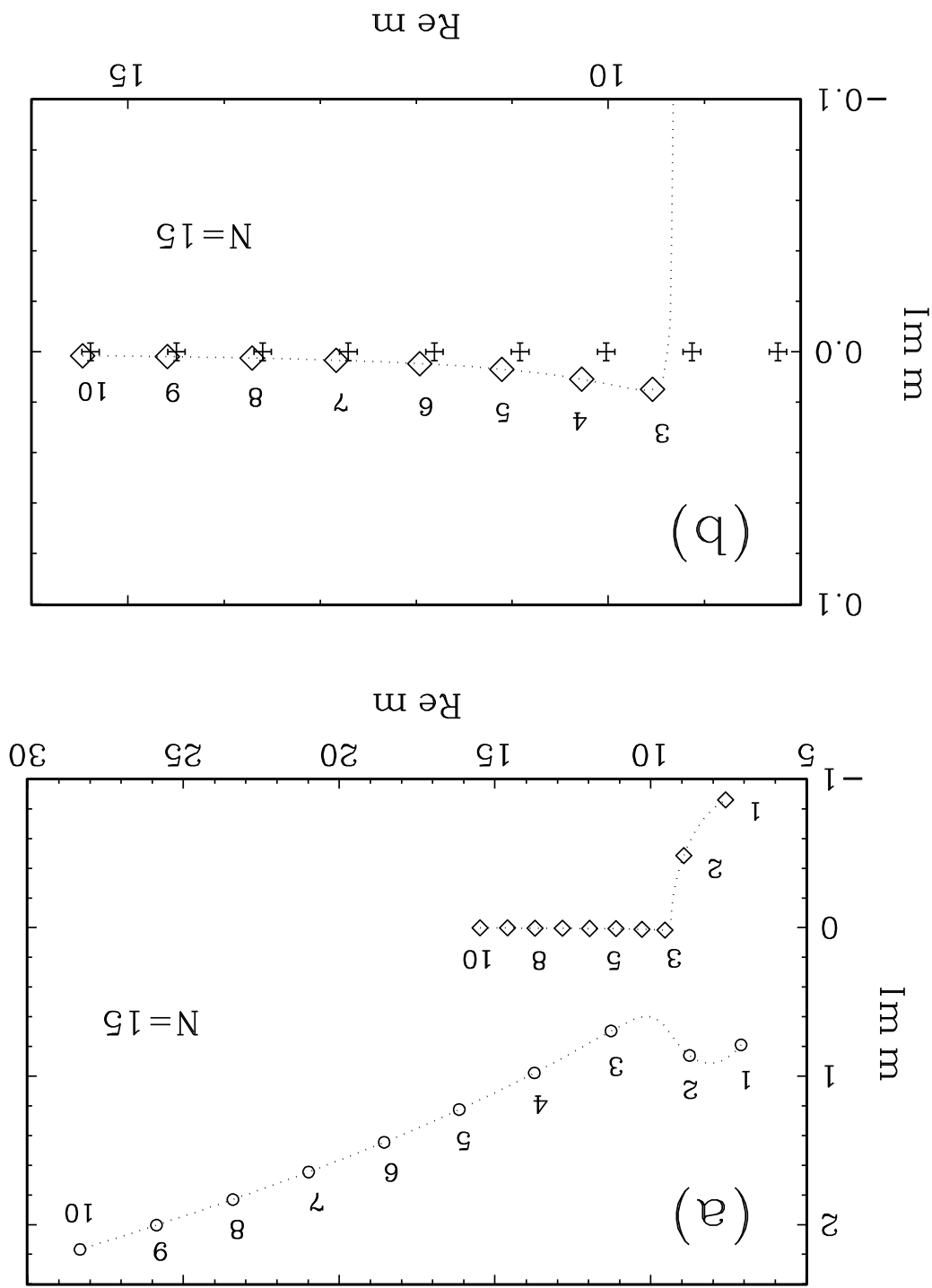
\begin{figure}
\caption{ \label{compind}
Trajectories of the saddle points $m_c(N)$ in the complex-$m$ 
plane for fixed harmonic order $N=15$ and varying
laser intensity $I = \zeta \times 10^{10}$ W/cm$^2$
(HG by H$^-$ ion in the laser field with the frequency 
$\omega = 0.0043$ is considered as in Fig.~\protect\ref{comp}).
The plot (a) gives general overview, the plot (b) shows vicinity
of the real-$m$ axis.
Positions of two saddle points for odd integer $N$ are denoted 
respectively by diamonds and circles with the numbers indicating
value of the factor $\zeta$, i.e. intensity in the units 10$^{10}$ W/cm$^2$. 
The crosses in the plot (b) show positions of the ATI threshold defined 
by formula \protect(\ref{mthrdef}) for the same values of $\zeta$.
}
\end{figure}

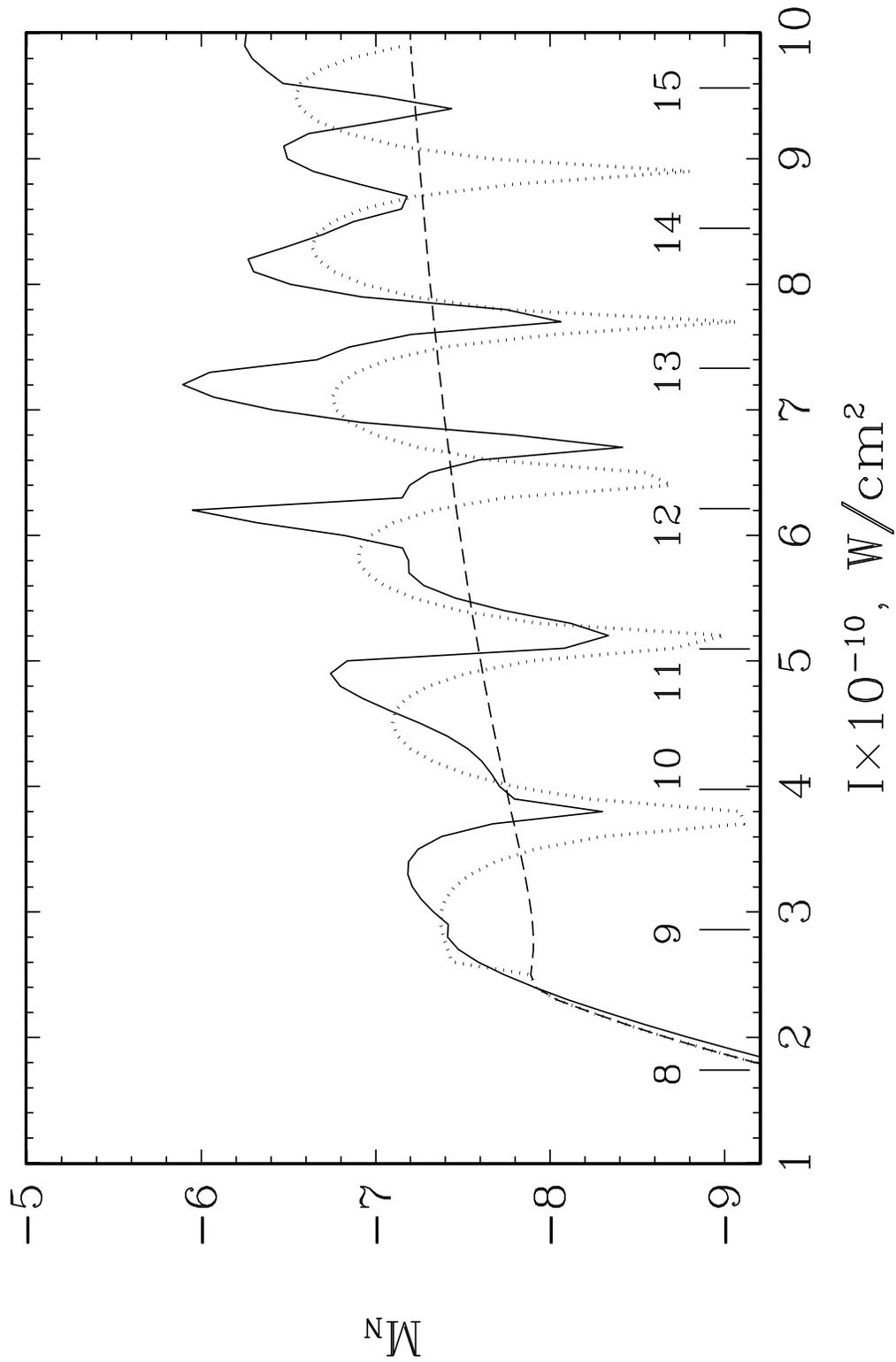
\begin{figure}
\caption{ \label{rateind}
The rate parameter $M_N \equiv 2 \log_{10} \left| d^+_N \right|$ 
for the harmonic $N=15$ as a function of the laser field 
intensity $I$ 
(HG by H$^-$ ion in the laser field with the frequency 
$\omega = 0.0043$).
Solid curve - our calculations \protect\cite{KOpap}
in the dipole-length gauge 
%using the expression for $B_{N m \, \mu}$ and 
performing numerical summation (\protect\ref{dfin})
over contributions of different ATI
channels, dashed curve -- present results within the saddle point 
approximation (\protect \ref{pssad}) for the summation over 
ATI channels contributions with a single saddle point $m_c(N)$
taken into account (namely, the saddle point shown by 
diamonds in Fig.~\protect\ref{compind}); 
dotted curve -- same but taking into account two saddle
points $m_c(N)$ (namely, these shown by 
diamonds and circles in Fig.~\protect\ref{comp}).
The bars with numbers $m$ indicate the threshold intensities
$I_m$ such that for $I>I_m$ the $m$-th ATI channel is closed
due to ponderomotive potential. 
}
\end{figure}

It is equally instructive to see how the rate of some individual 
harmonic depends on the laser intensity $I$. Fig.~\ref{compind}
shows trajectories of the saddle points as $I$ varies; only two
most important saddle points are considered. For small 
intensities only the saddle point shown by diamonds is operative
in evaluation of HG rate. This corresponds to the rapid-fall-off
regime beyond the plateau domain. Here ${\rm Im} \: m_c$ is large 
and negative. As $I$ increases, the saddle point moves to the real 
$m$ axis. After crossing the axis the trajectory abruptly bends
at $I$ about $3 \times 10^{10}$ W/cm$^2$. For higher intensities,
in the plateau domain, the second saddle point, shown by circles, 
should also be taken into account in calculation of HG rates. 
In this region the single-saddle-point approximation correctly 
reproduces an average behavior of the HG rate, whereas 
the two-saddle-point approximation shows also characteristic 
oscillations of the rate with $I$ (Fig.\ref{rateind}).
The positions of minima and maxima are correctly reproduced
for medium intensities. On the higher $I$ side an additional
maximum at $I = 9 \times 10^{10}$ W/cm$^2$ is beyond
the two-saddle-point approximation being tentatively
due to the contribution of the third saddle point omitted
in the present calculations. This contribution could be also 
the source of an additional minor structure in the medium 
intensity region. It could explain also the fact that the 
two-saddle-point approximation gives deeper minima in
the rates than the numerical summation in (\ref{dfin}).

The present calculations clearly show that the structures in
HG rates as function of $I$ is related to the interference effects,
in agreement with conclusion reached by Lewenstein {\it et al}
\cite{Lewphase}. The alternative explanation \cite{Brapid,B} 
relates this structure to the threshold
effects, namely to the successive closure of ATI channels by
increasing ponderomotive potential as $I$ grows.
However, the threshold effects should be manifested also
in the single-saddle-point approximation; Fig.~\ref{rateind}
demonstrates that this is not the case.

In the plateau regime the saddle point $m_c$ shown by diamonds 
in Fig.~\ref{rateind} lies in the complex-$m$ plane close to 
the threshold value $m_{\rm th}$ calculated for the
same laser intensity using formula (\ref{mthrdef}). 
As $I$ increases, $m_c$ tends to approach $m_{\rm th}$.
This behavior is in agreement with the three-step model 
developed by Corkum \cite{C} who presumed that the ATI 
electron leaves an atom with zero velocity. 
Note however, that this presumption is valid only within
single-saddle-point approximation (shown by diamonds in 
Fig.~\ref{compind}) and only in the plateau domain.
The point $m = m_{\rm th}$ is a branch point for any 
function which depends on $m$ via $K_m$. In particular,
generally the preexponential factor ${\cal Q}_{m \mu}$
(\ref{Q}) in expression (\ref{ps}) has such a branch 
point. In principle one can look for some refined
version of the saddle point approximation which accounts
for closeness of a saddle point and a branch point.
Note however that such modifications influence only
the preexponential factor but not the exponent which
is the principal object of interest. 
As discussed above, description of the transition between 
one-saddle point and two-saddle point regimes is more difficult 
for the semiclassical-type 
theory since it requires more sophisticated approach.
In the present case the situation is aggravated by the presence of
the branch point. It makes behavior of trajectories near
the 'collision point' quite different from the conventional 
cross-like pattern (Fig.~\ref{compind}) and leads to 
appearance of a spike on the dashed curve in Fig.~\ref{rateind}
at the borderline intensity $I = 2.7 \times 10^{10}$ W/cm$^2$.

%In our
%representation this presumption corresponds to 
%equality $m_c = m_{\rm th}$. It is to be compared with 
%the sa

%The model developed by Corkum \cite{C} presumes that
%the ATI electron leaves atom with zero velocity. In our
%representation this presumption corresponds to 
%equality $m_c = m_{\rm th}$. It is to be compared with 
%the saddle point trajectories shown in Fig.~\ref{comp}.

\section{CONCLUSION} \label{conc}

The present study has an objective to get a better insight into
the three-step mechanism of processes in strong laser field.
We take HG as the simplest case of three-step process 
and employ representation of its amplitude in terms of
amplitudes of physical, fully accomplished ATI and PLAR
processes. The number of alternative paths in the three-step
picture is generally infinite, each path being labeled by
the number $m$ 
%($m_{\rm th} < m < \infty$) 
of laser photons absorbed in the first step, ATI.
We explore the range of substantial intermediate ATI channels
and reveal crucial role of coherent interference. 
The effective scheme of summation of the contributions coming
from different intermediate ATI channels is developed
based on the saddle point method. Physically this approach 
is justified by the multiphoton nature of the process.
Due to it in the strong-field regime both 
ATI and PLAR amplitudes have phases that vary rapidly with $m$.
Within this framework we develop concept of effective ATI channels
that is important basically and useful for practical calculations. 
In this approach interference of infinite number of competing
paths is replaced by the single-path picture for generation
of high harmonic, or by interference of two paths for harmonics
lying in the plateau domain. In particular, the structure in HG 
rates in the latter domain is understood as the simple two-paths
interference pattern. 

The effective ATI channels are related to
the particular classical electron trajectories that ensure
that electron emitted from the atom at the first, ATI stage of
HG process returns to the core to make possible subsequent
laser assisted recombination. The possible trajectories possessing 
this property in principle comprise an infinite set differing
by the excursion time in continuum which is integer multiple
of laser period $T$. However due to quantum wave packet spread
only the trajectories with the shortest excursion time provide 
substantial contribution to HG rate. The crucial step in
this development is compexification of the problem: the initial 
and final time is complex-valued that makes entire classical
trajectory also complex-valued. The simplicity of description
achieved along these lines is in sharp contrast with the
approach based on the conventional real-time classical 
trajectories where the problems related to chaotic
irregular classical motion emerge in full scale \cite{Rost}.

\acknowledgements

This work has been supported by the Australian Research Council. 
V.~N.~O. acknowledges the hospitality of the staff of 
the School of Physics of UNSW where this work has been
carried out. 

\appendix

\section{ATI Amplitude in Adiabatic Approximation}
\label{A1}

The original expression for the ATI amplitude within the Keldysh \
\cite{Keldysh} approximation reads
\begin{eqnarray} \label{Adef}
A_{m}({\bf p}) =   
\frac{1}{T}\int \limits_{0}^{T} dt^\prime \: 
\langle \Phi_{{\bf p}}(t^\prime) \mid V_F(t) 
\hat{d}_{\mbox {\boldmath $\epsilon$}} \mid \Phi_a(t^\prime) 
\rangle ~,
\end{eqnarray}
where $\Phi_{\bf p}$ is the Volkov state of the electron with
the translational momentum ${\bf p}$, $\Phi_a$ is the electron
initial bound state. Details of all the definitions
could be found in Ref.~\cite{KOpap}; here we indicate only that
$V_F({\bf r}, t) = {\bf r} \cdot {\bf F} \cos \omega t $
describes an interaction of the active electron and the laser field, 
with $\omega$, ${\bf F}$ and $T$ being introduced in Sec.~\ref{threestep}.
The absolute value $p_m$ of the electron momentum in the $m$-the 
ATI channel is subject to energy conservation constraint being 
given by formula (\ref{Kmnm}) ($p_m=K_m$).

%The specific initial time $t^\prime = t_{m \mu}^\prime$ is defined
%by 
The adiabatic treatment of the ATI process was developed in 
Ref.~\cite{Multa} and subsequently applied in 
Refs.~\cite{KObi,KOlet,KOpap}. 
It presents the ATI amplitude $A_{m}({\bf p})$ as 
\begin{eqnarray} \label{A}
A_m^{(sp)}({\bf p}) & = & \sum_{\mu} \, A_{m \, \mu}^{(sp)}({\bf p}) ~, 
\\ \label{Amu}
A_{m \, \mu}^{(sp)}({\bf p}) 
& = & - \frac{(2 \pi)^2}{T} \, A_a \, \Gamma(1+\nu/2)
\, 2^{\nu/2} \, \kappa^\nu \, Y_{lm_{\rm az}}(\hat{{\bf p}}) \,
\frac{\exp \left[ i S(t^\prime_{m \mu}) \right]}
{\sqrt{- 2 \pi i S^{\prime \prime}(t^\prime_{m \mu})^{\nu+1} }} ~,
\end{eqnarray}
where $S(t)$ is the classical action 
\begin{eqnarray} \label{S}
S(t) = \frac{1}{2} \int^t d \tau \left({\bf p} + 
\frac{{\bf F}}{\omega} \sin \omega \tau \right)^2
-E_a t ~.
\end {eqnarray}
The position of the saddle points $t_{m \mu}^\prime$ in
the complex $t^\prime$ plane is defined by equation
\begin{eqnarray} \label{seqa}
S^\prime(t_{m \mu}^\prime) = 0 ~,
\end{eqnarray}
or, more explicitly,
\begin{eqnarray} \label{seqb}
\left( {\bf p} + \frac{{\bf F}}{\omega} 
\sin \omega t_{m \mu}^\prime \right)^2 + \kappa^2 = 0 ~.
\end{eqnarray}
Expression for HG amplitude (\ref{dfin}) refers to 
the particular saddle point ($\mu = \mu_0$) defined 
for the monochromatic laser field by the expressions
\begin{mathletters} \label{spATI}
\begin{eqnarray} 
\sin \omega t^\prime_{m \, \mu_0} & = & \frac{\omega}{F} 
\left(- K_m + i \kappa \right) ~, 
\\ 
\cos \omega t^\prime_{m \, \mu_0} & = & 
\sqrt{1 - \frac{\omega^2}{F^2} \left(- K_m + i \kappa \right)^2} ~. 
\end {eqnarray}
\end{mathletters}

The other notations are as follows: $\frac{1}{2} \kappa^2$ is
initial electron binding energy, $\nu = Z/\kappa$, $Z$
is the charge of the atomic residual core ($\nu=Z=0$ for a
negative ion), $l$, $m_{\rm az}$ are the active electron orbital 
momentum and its projection in the initial state. 
In the present context the unit vector $\hat{{\bf p}}$
coincides with $\hat{{\bf F}} = {\bf F}/F$.
The coefficients $A_a$ are specified in Ref.~\cite{Multa} being
tabulated for many atoms and ions \cite{RS}.
Mathematically formula (\ref{A}) is obtained by using the saddle 
point method to carry out integration over time variable $t^\prime$
in formula (\ref{Adef}). The saddle point positions are defined by
Eqs.~(\ref{seqa}) or (\ref{seqb}) that are to be considered 
together with the energy conservation constraint (\ref{Kmnm}).
The summation in (\ref{A}) runs over the saddle points
$t^\prime_{m \, \mu}$ in the plane of the complex-valued 
time $t^\prime$. The saddle points $t^\prime_{m \mu}$ lie 
symmetrically with respect to the real $t^\prime$ axis. 
For the monochromatic laser field there are four saddle 
points in the interval 
$ 0 \leq {\rm Re} \, t^\prime_{m \mu} \leq T$, two of them lying 
in the upper half plane (${\rm Im} \, t^\prime_{m \mu} > 0$).
Only these two saddle points are included into the summation in (\ref{A}). 
If ATI is considered as the first step  in HG process, then,
as discussed in detail in Ref.~\cite{KOpap}, only one of 
these two saddle points is effectively operative, namely,
that specified by formulae (\ref{spATI}).

\section{LAR and PLAR Amplitudes in Adiabatic Approximation}
\label{A2}
\label{LAR}

The PLAR amplitude defined as 
\begin{eqnarray} \label{B}
B_{N \, m \mu_0}({\bf p}) & = & - \,  
\frac{1}{2 \pi T}\int \limits_{0}^{T} dt \: 
\frac{1}{ R_0(t,t_{m \mu_0}^\prime) } \,
\langle \Phi_a(t) \mid \exp ( i \Omega t ) \, 
\hat{d}_{\mbox {\boldmath $\epsilon$}} \mid 
\Phi_{{\bf p}} (t) \rangle ~,
\end{eqnarray}
differs from the LAR (Laser Assisted Recombination)
amplitude $C_{m}({\bf p})$
\begin{eqnarray} \label{rec}
C_{N \, m}({\bf p}) = \frac{1}{2 \pi T}\int \limits_{0}^{T} dt \,
\langle \Phi_{\bf p} (t) \mid \exp ( i \Omega t ) \, 
\hat{d}_{\mbox {\boldmath $\epsilon$}} \mid
\Phi_a (t) \rangle 
\end{eqnarray}
only by the ``propagation'' or ``expansion'' factor 
$1/\left[R_0(t,t_{m \mu_0}^\prime) \right]$ in the integrand.
The distance passed by electron in the laser field
between initial time $t^\prime$ and final time $t$
is approximated as \cite{KOpap}
\begin{eqnarray} \label{prop1}
R_0(t,t^\prime) = \frac{F}{\omega^2} \left( \cos \omega t - 
\cos \omega t^\prime \right) ~.
\end{eqnarray}
In (\ref{B}), (\ref{rec}) $\hat{d}_{\mbox {\boldmath $\epsilon$}} =  
{\mbox {\boldmath $\epsilon$}} \cdot {\bf r}$ 
is the dipole interaction operator.

Consider electron in the laser-dressed continuum state with
the translational momentum ${\bf p}$.
It can recombine to the bound state $\Phi_a$ with emission of 
the photon. Possible frequencies of the emitted photon 
$\Omega^{\rm LAR}_j$ form an equidistant pattern:
\begin{mathletters} \label{spectrum}
\begin{eqnarray} \label{omLAR}
\Omega^{\rm LAR}_M = \frac{1}{2} \, {\bf p}^2 + U_p
- E_a + M \omega 
\end{eqnarray}
with an integer $M$. Expression (\ref{omLAR})
can be reparametrized to the form
\begin{eqnarray}
\Omega_j^{\rm LAR} = (J + \eta) \omega ~,
\end{eqnarray}
\end{mathletters}
where $J$ is another integer and the fractional parameter 
$\eta$ ($0 \leq \eta < 1$) is governed by the value of 
the initial momentum $p$.
In the zero-laser-field limit ($F \rightarrow 0$) only emission 
of the photon with the frequency 
$\Omega^{\rm LAR}_{F \rightarrow 0} = \frac{1}{2} \, {\bf p}^2 -E_a$
is allowed. The presence of intensive laser field makes possible 
the processes when laser photons are absorbed from the field or 
transmitted to it, and thus the entire spectrum (\ref{spectrum}) 
is produced. Such a LAR process has not yet received much attention 
in literature; as far as we know, our recent study \cite{LAR} 
is the only theoretical paper on the subject. In it
the initial electron translational momentum ${\bf p}$ was considered 
as an arbitrary input parameter as required in applications
to LAR process in laser plasma. In other terms,  
the fractional parameter $\eta$ was arbitrary.

Here we are interested in LAR as a constituent part of HG process.
This application has some particular features.
First, all the frequencies in the photon spectrum are integer
multiples of $\omega$, i.e. $\eta = 0$. 
This happens because the initial translational momentum $p$ of LAR
process could not be arbitrary since only the discrete subset
of continuum states is populated by ATI from the bound state.
Namely, the translational momentum is subject to the constraint
(\ref{Kmnm}) with $m$ being the number of laser photons absorbed 
on the initial ATI stage of HG process. 

Second, when an individual harmonic is considered, its
order $N$ is fixed, and the parameter $m$ is scanned when
summation is carried out according 
to the expression (\ref{sumc}). Equivalently, the electron 
momentum $p = K_m$, is scanned along the discrete set of allowed 
values $K_m$, see Eq.~(\ref{Kmnm}). This is in variance
with the laser plasma applications when it is natural to presume
that ${\bf p}$ is fixed and $\Omega^{\rm LAR}$ is scanned. 
The third special feature is that the electron translational
momentum ${\bf p} = {\bf K}_m$ is parallel to the electric field 
amplitude ${\bf F}$. 

Finally, the object of interest in HG theory is the PLAR 
amplitude (\ref{B}) rather than LAR amplitude (\ref{rec}),
although both amplitudes are quite similar.
Below we briefly expose modification of some results 
in the LAR theory relevant to the present application.

The expressions (\ref{B}) and (\ref{rec}), respectively, for 
PLAR and LAR process amplitudes are valid in the Keldysh-type 
approximation. By using the Fourier transformation they are 
rewritten as
\begin{eqnarray} 
\label{BF}
B_{N \, m \mu_0}({\bf K}_m) & = & - \, 
\frac{1}{2 \pi T} \, \int \limits_{0}^{T} dt \:
\frac{1}{ R_0(t,t_{m \mu_0}^\prime) } \,
\exp \left\{ i \left[\Omega t - S(t) \right] \right\} \:
\tilde{\phi}_a^{(\epsilon)} \left( - {\bf K}_m - 
\frac{{\bf F}}{\omega} \sin \omega t \right) ~,
\\
\label{CF} 
C_{N m}({\bf K}_m) & = & - \, \frac{1}{2 \pi T} \, \int \limits_{0}^{T} dt \:
\exp \left\{ i \left[\Omega t - S(t) \right] \right\} \:
\tilde{\phi}_a^{(\epsilon)} \left( - {\bf K}_m - 
\frac{{\bf F}}{\omega} \sin \omega t \right) ~,
\end{eqnarray}
where the classical action $S(t)$ is introduced above by formula (\ref{S}).
The function $\tilde{\phi}_a^{(\epsilon)}({\bf q})$ is defined as
\begin{eqnarray} 
\tilde{\phi}_a^{(\epsilon)}({\bf q}) = i 
\left( {\mbox {\boldmath $\epsilon$}} 
\cdot \nabla_{\bf q} \right) \tilde{\phi}_a({\bf q}) ~. 
\end{eqnarray}
where $\tilde{\phi}_a({\bf q})$ is the Fourier transform of 
the bound state wave function $\phi_a({\bf r})$:  
\begin{eqnarray} \label{wF}
\tilde{\phi}_a({\bf q}) = \int d^3 {\bf r} \, 
\exp( - i {\bf q} {\bf r} ) \, \phi_a({\bf r}) ~ .
\end {eqnarray}

The time integrals in (\ref{BF}) and (\ref{CF}) could be 
evaluated using the saddle point approximation:
\begin{eqnarray} \label{Bsprep}
B_{N \, m \, \mu_0} & = &
\frac{1}{2 \pi T} \: \sum_\mu \:
\frac{\omega^2}
{F \left(\cos \omega t^\prime_{m \, \mu_0}  - 
\cos \omega t_{m \, \mu} \right) } \:
\tilde{\phi}_a^{(\epsilon)} \left( - {\bf K}_m - 
\frac{{\bf F}}{\omega} \sin \omega t_{m \mu} \right)
\times \nonumber \\ & &\times
\sqrt{\frac{2 \pi}{ i S^{\prime \prime} \left(t_{m \mu} \right)}} \:
\exp \left\{ i \left[N \omega t_{m \mu} - S(t_{m \mu}) \right] \right\} ~.
%\end{eqnarray}
%\begin{eqnarray} 
\\
\label{Csp}
C_{N \, m} & = & -
\frac{1}{2 \pi T} \: \sum_\mu \: 
\tilde{\phi}_a^{(\epsilon)} \left( - {\bf K}_m - 
\frac{{\bf F}}{\omega} \sin \omega t_{m \mu} \right)
\times \nonumber \\ & &\times
\:
\sqrt{\frac{2 \pi}{ i S^{\prime \prime} \left(t_{m \mu} \right)}} \:
\exp \left\{ i \left[N \omega \, t_{m \mu} - S(t_{m \mu}) \right] \right\} ~,
\end{eqnarray}
where summation is to be taken over the saddle points $t_{m \mu}$
operative in the contour integration.
The position of saddle points 
in the complex $t$-plane is governed by the equation 
\begin{eqnarray} \label{spt}
S^\prime(t_{m \mu}) - \Omega = 0 ~,
\end{eqnarray}  
or, more explicitly
\begin{eqnarray} \label{sptexp}
\frac{1}{2} \left({\bf p} +
\frac{{\bf F}}{\omega} \sin \omega t_{m \mu} \right)^2 =
%- \frac{1}{2} \kappa^2 
E_a + \Omega ~.
\end{eqnarray}
Its solution is 
\begin{eqnarray} \label{sprad}
\sin \omega t_{m \, \mu} = \frac{\omega}{F} 
\left(- K_m \pm \sqrt{2 N \omega - \kappa^2 } \right) ~. 
\end {eqnarray}
For $N=0$ the formula (\ref{sprad}) coincides with that governing the saddle
point position in case of ATI processes [see Eq.~(\ref{spATI})]. 
As in the latter case,
Eq.~(\ref{spt}) has four solution per the field cycle 
(i.e for $0 < {\rm Re} \, t_{m \mu} < T$). 
Physically $t^\prime_{m \, \mu_0}$ is the time when the electron
is emitted at the first step of HG process and $t_{m \, \mu}$ 
is the time of electron return back to the initial 
bound state at the last stage. Note however that both
$t^\prime_{m \, \mu_0}$ and $t_{m \, \mu}$ are generally 
complex-valued.
%The formula (\ref{Csp}) is rewritten as
%\begin{eqnarray} 
%\label{Csprep}
%C_{N \, m} & = & -
%\frac{1}{2 \pi T} \: \sum_\mu \: 
%\tilde{\Phi}_a^{(\epsilon)} \left( - {\bf K}_m - {\bf k}_{t_{m \mu}} \right)
%%\times \nonumber \\ & &\times
%\:
%\sqrt{\frac{2 \pi}{ i S^{\prime \prime} \left(t_{m \mu} \right)}} \:
%\exp \left\{ i \left[N \omega t_{m \mu} - S(t_{m \mu}) \right] \right\} ~,
%\end{eqnarray}

Formula (\ref{sprad}) might give real-valued saddle points 
$t_{m \mu}$ that correspond to the classically allowed LAR or PLAR. 
According to Eq.~(\ref{sptexp}) this means emission 
of the high-energy photon at the real moments of time when 
the instantaneous kinetic energy of the electron in the laser field 
differs from the bound state energy 
%$- \frac{1}{2} \kappa^2$ 
$E_a$ by $\Omega$. 
In this respect LAR or PLAR processes differ basically from ATI 
process which is always described by the complex-valued saddle 
points that corresponds to the classically forbidden or tunneling
transitions.
%Summation over $\mu$ in (\ref{Csp}) or 
%(\ref{Bsprep}) is to be limited to these points. 
The classically allowed transitions in LAR or PLAR are absent for small
$N$ when $N < \kappa^2/(2 \omega)$. However we are interested 
only in generation of harmonics with sufficiently high order: 
\begin{eqnarray} \label{Ng}
N > \frac{\kappa^2}{2 \omega}  ~. 
\end{eqnarray}
In this case the classically allowed population is operative 
provided $m$ is not too large, namely
\begin{mathletters}
\begin{eqnarray}
K_m < \frac{F}{\omega} + \sqrt{2 N \omega - \kappa^2} ~,
\end{eqnarray} 
i.e.
\begin{eqnarray}
m < m_N^+ \equiv N  + 3 \frac{U_p}{\omega} + 
\frac{F}{\omega^2} \: \sqrt{2 N \omega - \kappa^2} ~.
\end{eqnarray} 
\end{mathletters}
If $N$ is sufficiently large, namely if
\begin{mathletters}
\begin{eqnarray}
\frac{F}{\omega} <  \sqrt{2 N \omega - \kappa^2} ~,
\end{eqnarray}
i.e.
\begin{eqnarray} \label{Nlarge}
N > \frac{\kappa^2}{2 \omega} + \frac{2 U_p}{\omega} ~,
\end{eqnarray}
\end{mathletters}
then the domain of classically populated $m$-channels is bounded
from below by the condition
\begin{mathletters}
\begin{eqnarray}
K_m > - \frac{F}{\omega} + \sqrt{2 N \omega - \kappa^2} ~,
\end{eqnarray}
i.e. 
\begin{eqnarray}
m > m_N^- \equiv N  + \frac{3 U_p}{\omega} - 
\frac{F}{\omega^2} \: \sqrt{2 N \omega - \kappa^2} ~. 
\end{eqnarray}
\end{mathletters}
In the interval $m_N^- < m < m_N^+$ the classical population
is governed by two real saddle points $t_{m \mu}$ per field cycle.
Note that for the real LAR or PLAR process one always implies that 
the $m$-th ATI channel is open, i.e. $m$ is sufficiently large:
\begin{eqnarray} \label{mthr}
m > m_{\rm th}   
\end{eqnarray}
where $m_{\rm th}$ is defined by Eq.~(\ref{mthrdef}).
The same constraint is assumed also in the sum over $m$ 
in formula (\ref{sumc}) implementing the three-step
mechanism of HG \cite{KOlet,KOpap}.

For intermediate $N$ lying in the interval
\begin{eqnarray} \label{Nint}
\frac{\kappa^2}{2 \omega} < N < \frac{\kappa^2}{2 \omega} 
+ \frac{2 U_p}{\omega}  
\end{eqnarray}
another regime of classical population becomes possible.
Namely, if the condition
\begin{mathletters}
\begin{eqnarray}
K_m < \frac{F}{\omega} - \sqrt{2 N \omega - \kappa^2} ~,
\end{eqnarray} 
i.e.
\begin{eqnarray}
m_{\rm th} < m < m_N^- \equiv N  + \frac{3 U_p}{\omega} - 
\frac{F}{\omega^2} \: \sqrt{2 N \omega - \kappa^2} 
\end{eqnarray}
\end{mathletters} 
is satisfied, there are four saddle points $t_{m \mu}$ per 
field cycle.

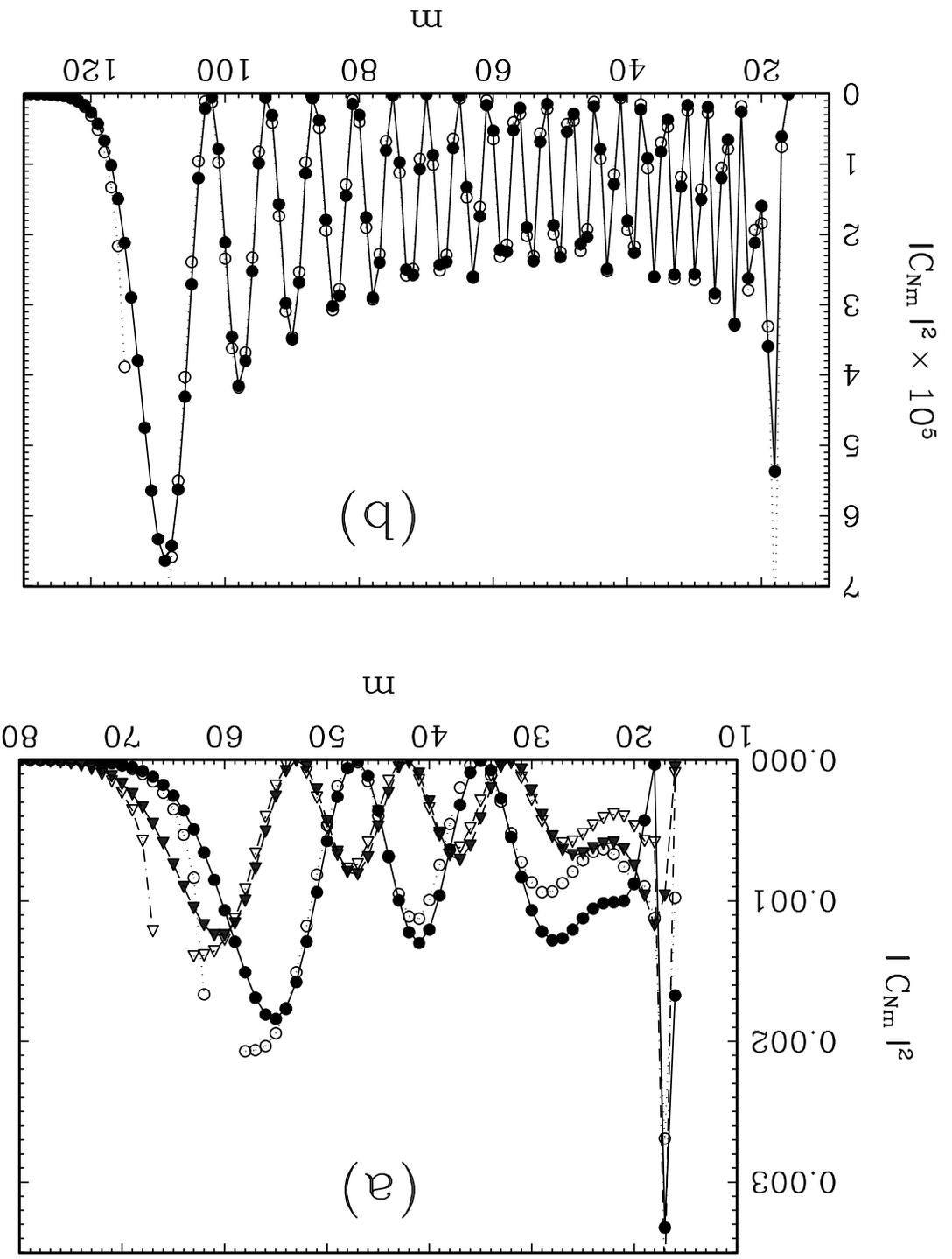
\begin{figure}
\caption{ \label{C}
Squared modulus of the amplitude $C_{Nm}$
of the laser assisted recombination 
of electron into H$^-$ bound state in the laser field 
with the frequency $\omega = 0.0043$ and intensity 
$I = 10^{11}$ W/cm$^2$. The harmonic order is fixed
[$N=13$ (circles) and $N=15$ (triangles) in (a)
and $N=39$ (circles) in (b)].
Closed symbols -- calculations using the formula (\protect \ref{CF})
with numerical integration over $t$; open symbols --
adiabatic approximation (\protect \ref{Csp}) with various number 
of saddle points taken into account for different $m$ as
described in the text. 
}
\end{figure}

In Fig.~\ref{C} we illustrate these results taking LAR of
electron on hydrogen atom with formation of negative H$^-$
ion. The laser field intensity is $I = 10^{11}$ W/cm$^2$. 
Open ATI channels lie at $m \geq 16$.
The intermediate $N$ regime (\ref{Nint}) 
takes place for $7 \leq N \leq 24$. In Fig. \ref{C}a we
show $m$ distributions of LAR modulus squared amplitudes
$|C_{N m}|^2$ for two adjacent $N$ from the intermediate-$N$
domain, $N=13$ and $N=15$. For $N=13$ the classically forbidden
domain is $m \geq 62$; the classical population 
in the two-saddle-point regime is operative for $19 \leq m \leq 61$,
the four-saddle-point regime is operative for $16 \leq m \leq 18$.  
For $N=15$ the classical population is forbidden for
$m \geq 66$; it is allowed in the two-saddle-point regime 
for $18 \leq m \leq 66$; the four-saddle-point regime is 
operative for $16 \leq m \leq 17$.  
The region of transition from classically allowed to classically
forbidden LAR requires some more elaborate treatment to 
reproduce related transition pattern similar to Airy function.
The special analysis is required also to describe
a transition from two-saddle-point to four-saddle point regime.
However here we do not pursue the objective of detailed
uniform description of LAR or PLAR amplitudes.

Note that for $N=0$ the saddle points coincide with the
singularity of the integrand $\tilde{\phi}_a^{(\epsilon)}({\bf q})$.
This coincidence occurs also in the adiabatic treatment of
ATI. The difference is that in the particular case of above 
threshold detachment of negative ion the singularity in
the integrand is canceled and the standard form of the saddle
point approximation is applicable. Otherwise (i.e. in case 
of non-zero atomic core charge) the saddle point method is 
to be applied in the somewhat modified form.
In case of LAR the singularity is always present, and hence
the modified form of the saddle point method should be applied
for $N=0$. Respectively, the small $N$ case requires 
some special treatment (some sort of uniform approximation;
note that all these modifications affect only the pre-exponential 
factor, but the principal exponent remains the same).
Below we are interested in the large $N$ case that allows us 
to put aside these complications. However, it is to be 
remembered that coincidence of the saddle point and
the singularity of the momentum space wave function means 
that only large-$r$ asymptote of the initial state
wave function $\varphi_a({\bf r})$ in the coordinate 
space is of importance. This is the case for ATI, but 
not for LAR or PLAR, as follows from the preceding discussion.
In other words, the LAR amplitude (or its PLAR counterpart)
generally are more sensitive for the short-range behavior 
of the wave function.

\begin{references}
\bibitem{ftnt1}
In the detailed picture \cite{Multa} 
an electron emerges from the under barrier at the
source point separated from the core. There are
two such points lying up and down the electric field
vector in the laser wave. The related contributions
interfere giving raise to the oscillatory patterns
in the ATI angular distributions. 
%The high energy photon
%is emitted when the electron returns to the core.
These physically essential features of HG mechanism 
are not manifested in the saddle point equation (\ref{z0})
governing the momentum $K_{m_c}$: this equation does
not distinguish between position in space of the 
source points and the atomic core.
\end{references}
\end{document}